\documentclass[aps,prx,twocolumn,floatfix,superscriptaddress,nofootinbib]{revtex4-2}
\usepackage{amsmath,amssymb,amsthm,mathtools,graphicx,booktabs,enumitem}
\usepackage{algpseudocode}
\usepackage{newfloat}
\DeclareFloatingEnvironment[fileext=loa,name=Algorithm,placement=t]{algorithm}
\usepackage{hyperref}
\setlist{nosep}
\newtheorem{theorem}{Theorem}
\newtheorem{lemma}[theorem]{Lemma}
\newtheorem{proposition}[theorem]{Proposition}
\newtheorem{corollary}[theorem]{Corollary}
\theoremstyle{remark}\theoremstyle{plain}
\newcommand{\F}{\mathbb F_2}
\newcommand{\E}{\mathbb E}
\newcommand{\TV}{\operatorname{TV}}
\newcommand{\lrw}{\operatorname{lrw}}
\newcommand{\SR}{\operatorname{SR}}
\newcommand{\CZ}{\operatorname{CZ}}
\newcommand{\CNOT}{\operatorname{CNOT}}
\newcommand{\rank}{\operatorname{rank}_{\F}}
\newcommand{\kl}{\operatorname{kl}}
\newcommand{\ket}[1]{|#1\rangle}
\newcommand{\bra}[1]{\langle#1|}
\newcommand{\ip}[2]{\langle#1,#2\rangle}
\DeclareMathOperator{\spanF}{span}

\begin{document}

\title{Classical sampling of verifiable graph-state quantum advantage experiments}

\author{Zhongpu Qiu}
\email{slszqzp@outlook.com}
\affiliation{Shanghai Artificial Intelligence Laboratory, Shanghai, China}
\affiliation{Beijing Normal University, Beijing, China}
\author{Yuqi Qing}
\affiliation{Tsung-Dao Lee Institute, Shanghai Jiao Tong University, Shanghai, 201210, China}
\author{Jielan Li}
\affiliation{Shanghai Artificial Intelligence Laboratory, Shanghai, China}
\author{Hongxia Hao}
\email{haohongxia@pjlab.org.cn}
\affiliation{Shanghai Artificial Intelligence Laboratory, Shanghai, China}

\begin{abstract}
Quantum computers are expected to outperform classical ones on some tasks, but a convincing demonstration must be both hard to reproduce classically and verifiable. A recent proposal achieves verifiability by preparing graph states with built-in error detection, so that the quality of the experiment can be certified without simulating it; hardness is added by measuring in a non-Clifford basis or by inserting T gates. We show that the two instances posted to demonstrate this proposal on IBM hardware can be sampled classically on eight graphics processing units, with higher quality than the hardware in comparable or shorter time. For the 70-qubit instance our samples are drawn from the ideal output distribution; for the 64-qubit instance with 314 T gates they come with a certified fidelity of at least 0.908, against 0.349 for the hardware. The reason lies in verification itself: T gates can only be placed where the error detection cannot see them, which pushes them towards the end of the circuit, where they add little classical cost. Our results identify the features that future verifiable advantage experiments must control.
\end{abstract}

\maketitle

\section{Introduction}

Quantum computers promise to solve problems that are out of reach for classical machines, from the simulation of quantum matter~\cite{Feynman1982} to the factoring of large integers~\cite{Shor1997}. These applications require fault-tolerant devices that do not yet exist. In the meantime, a more modest goal has organized much of the field: to show, on present noisy processors, a single well-defined computation that no classical computer can reproduce in a reasonable time~\cite{Preskill2018}. Such a demonstration of quantum advantage would establish that the exponential state space of a quantum device is accessible in practice.

Sampling the output of random quantum circuits became the leading candidate for this demonstration~\cite{Boixo2018,Bouland2019,Hangleiter2023}. In 2019 a 53-qubit superconducting processor produced samples in 200~s for which the best available classical estimate was 10,000 years~\cite{Arute2019}, and larger circuits~\cite{Wu2021,Morvan2024} and photonic experiments~\cite{Zhong2020} followed. The classical side did not stand still. The estimate of 10,000 years referred to one particular algorithm, not to a lower bound. Soon after, about two and a half days on a supercomputer were estimated for the same task~\cite{Pednault2019}. Tensor-network contraction~\cite{markov,GrayKourtis2021} then reduced the actual cost to days on a small cluster of graphics processing units (GPUs)~\cite{PanZhang2022}, hours~\cite{Pan2022} and minutes~\cite{Liu2021SC}, and finally produced uncorrelated samples of higher quality in less time than the experiment~\cite{Zhao2025}. Classical algorithms also reproduced the photonic experiments~\cite{Oh2024} and, beyond sampling, the expectation values measured in a 127-qubit simulation of spin dynamics~\cite{Kim2023,Tindall2024,Begusic2024}. Each of these exchanges moved the frontier and sharpened the understanding of what makes a quantum circuit hard.

They also exposed a structural weakness of the experiments. A quantum advantage experiment has to meet two requirements that pull in opposite directions: the task should be hard for every classical method, and the experimenter should be able to show that the quantum processor performed it. A circuit that no classical computer can simulate is also a circuit whose output no classical computer can check. In random circuit sampling the fidelity is estimated by linear cross-entropy benchmarking (XEB), which scores the samples with ideal output probabilities; for the hardest circuits these cannot be computed, so the estimate is extrapolated from circuits that can still be simulated. The benchmark is also fragile: high XEB values can be reached without sampling from the ideal distribution~\cite{Barak2021,Gao2024}, and its hardness rests on conjectures that fail for shallow circuits~\cite{AaronsonGunn2020}. Schemes with built-in verification exist, for example for instantaneous quantum polynomial-time circuits, in which a secret structure hidden in the circuit lets the verifier check the samples~\cite{ShepherdBremner2009,Bremner2025}. Their security rests on the assumption that this structure cannot be extracted classically, and the first such scheme was broken in exactly this way~\cite{KahanamokuMeyer2023}. Other recent approaches use circuits whose output is concentrated on a single bit string that can be checked~\cite{AaronsonZhang2024}, certify randomness with verification on classical supercomputers~\cite{Liu2025certified}, or measure expectation values instead of samples~\cite{Google2025echoes}.

\begin{figure*}[tb]
\includegraphics[width=\textwidth]{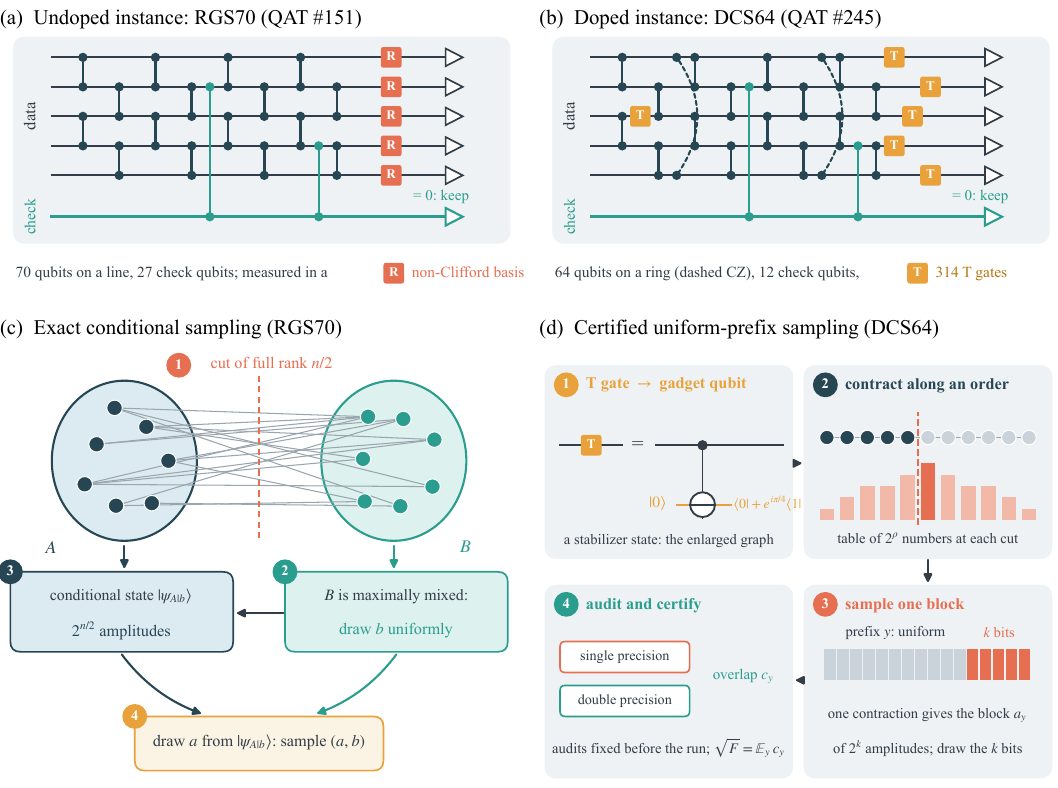}
\caption{The two instances and the classical samplers developed for them. (a) The undoped instance RGS70 (QAT~\#151): a random Clifford circuit prepares a graph state on 70 qubits on a line, check qubits detect faults, only shots with zero syndrome are kept, and the output is measured in a non-Clifford basis (R). (b) The doped instance DCS64 (QAT~\#245): the same construction on 64 qubits on a ring, with T gates inserted only at allowed positions. (c) Exact conditional sampling for RGS70 (Algorithm~\ref{alg:undoped}): across a balanced cut of full rank the half $B$ is maximally mixed, so its bits $b$ are drawn uniformly and the bits $a$ of the other half from the conditional state $|\psi_{A|b}\rangle$. (d) Certified uniform-prefix sampling for DCS64 (Algorithm~\ref{alg:doped}): each T gate is replaced by a gadget qubit, the enlarged graph is contracted along a linear order with the last $k$ output bits left open, each sample draws a uniform prefix $y$ and its remaining bits exactly from the block $a_y$, and audits fixed before the run recompute the block in double precision to certify the fidelity. Circuits are schematic.}
\label{fig:concept}
\end{figure*}

Martiel \emph{et al.}~\cite{Martiel2026} proposed a different way out (Fig.~\ref{fig:concept}). A circuit made of Clifford gates, which a classical computer can simulate efficiently, prepares a random graph state. A few additional check qubits detect faults during the run~\cite{Martiel2025,delfosse,paulice}, and only the shots in which no check fires are kept. Because the circuit is Clifford, the fidelity of the kept shots can be measured directly~\cite{FlammiaLiu2011,daSilva2011}. Hardness is then added in a way that leaves this measurement valid: either by measuring the output in a non-Clifford basis, or by inserting T gates only where they do not disturb the checks, so that the fidelity bound of the Clifford circuit carries over. The approach promises what random circuit sampling lacks: a certificate of quality that does not require simulating the hard circuit. It also attracted wide attention. On 30 July 2026, IBM and the University of Chicago announced that a doped instance, with 70 data qubits, 2,415 two-qubit gates and 468 T gates, run in about 15 minutes, met the criteria for quantum advantage; the announcement stated that the field was now ``firmly in the quantum advantage era''~\cite{IBMPress2026}. The circuits, samples and claims were posted on the Quantum Advantage Tracker (QAT), a public platform on which quantum and classical results for the same instances are compared and checked~\cite{QAT}.

The classical response followed the familiar pattern. The first doped instance (QAT~\#228) placed its 70 qubits on a line, and the exact probabilities of all its hardware samples were soon computed classically by contracting along the time direction, which the open boundary makes cheap~\cite{Manabe2026,QAT}. The experimenters then moved the qubits onto a ring and posted a 64-qubit instance with 314 T gates, the one reported in their paper~\cite{Martiel2026}. Together with an earlier undoped instance measured in a non-Clifford basis, this left two instances standing when we began: the undoped 70-qubit instance, which we call RGS70 (QAT~\#151) and for which the experimenters estimated classical costs of up to $10^{25}$~s~\cite{vidal,Bravyi2019}, and the 64-qubit doped instance, which we call DCS64 (QAT~\#245).

Here we ask what the certificate costs in classical hardness. Past exchanges have shown that a single improved algorithm can settle a particular instance; the more useful outcome is an understanding of which features of a circuit make it hard. For verifiable experiments the question has a specific form: does the structure that makes the fidelity measurable also limit how hard the circuit can be?

Our answer has three parts. First, both instances can be sampled classically, on eight GPUs, with higher quality than the experiments and in comparable or shorter time (Table~\ref{tab:results}). For RGS70 the samples are drawn from the ideal output distribution, up to floating-point rounding. Its graph state is so strongly entangled that half of the output bits are fair coin flips, and the other half follow from a conditional state that fits into the memory of two GPUs (Sec.~\ref{sec:exact}). For DCS64 the sampler is approximate but, like the experiment, comes with a certificate: from audits fixed before the run we show that the samples are drawn from a state whose fidelity with the ideal state is at least 0.908, against a certified fidelity of 0.349 for the hardware. The two numbers are defined differently but play the same role: each bounds how far the produced samples can be from ideal ones (Sec.~\ref{sec:doped}).

Second, the reason is tied to verifiability (Sec.~\ref{sec:mechanism}). A T gate leaves the checks intact only where a $Z$ fault at the same place would go undetected~\cite{Martiel2025}; we call such places allowed positions. Good checks leave few allowed positions, and fewest early in the circuit; in DCS64 only one of the 314 T gates lies in the first 50 of 73 layers [Fig.~\ref{fig:mech}(a)]. Late T gates, however, add little classical cost. What sets the cost of our method is the early axis rank, the number of independent T gates early in the circuit, and the early entanglement, the entanglement created by the Clifford circuit before them; the total number of T gates does not enter. Random checked circuits confirm that the early axis rank collapses once the circuit is deep enough for the checks to cover its early region. In this regime verification and hardness compete for the same resource.

Third, the analysis shows where this picture stops. Counterexamples rule out the simpler statements one might hope for: neither the number of checks nor the number of allowed positions determines the classical cost, and many allowed positions do not force a weaker certificate. The mechanism is conditional, not a universal trade-off, and it tells future experiments which quantities they must raise.

The tools we use are known: rank-width~\cite{OumSeymour2006,oum} methods for graph states~\cite{VandenNest2007,Kuyanov2026}, T-gate injection as applied to DCS64 in Ref.~\cite{Renaud249}, uniform-prefix sampling~\cite{Manabe2026} and the compatibility rule~\cite{Martiel2025}. What is new is their combination into an exact sampler for undoped and a certified sampler for doped instances, the two classical data sets with public verifiers, and the structural analysis that explains why these instances are tractable. As in the earlier exchanges on random circuit sampling, all samples and code are public. Concurrent work samples RGS70 exactly on CPUs at longer run time~\cite{QAT262}, and Ref.~\cite{Renaud249} computed exact probabilities of the DCS64 hardware samples.

\section{Verifiable graph-state sampling}
\label{sec:setting}

\paragraph*{Graph states and cut rank.} For a graph $G$ on $n$ vertices with adjacency matrix $\Gamma$ over $\F$, the graph state is $\ket G=2^{-n/2}\sum_x(-1)^{\sum_{i<j}\Gamma_{ij}x_ix_j}\ket x$. Every stabilizer state is a graph state up to single-qubit Clifford gates~\cite{VandenNest2004}. For a bipartition $A\cup B$, the reduced state on either side has a flat spectrum of $2^{\rho}$ nonzero eigenvalues, where $\rho=\rank\Gamma_{AB}$ is the cut rank~\cite{Hein2004,VandenNest2007}. The cut rank is the entanglement entropy in bits, and it sets the cost of every method in this paper.

\paragraph*{The experiments.} Both instances apply alternating layers of CZ gates and random single-qubit Clifford gates to qubits on a line or a ring, with check qubits that measure the spacetime checks, and keep the shots in which no check fires [Fig.~\ref{fig:concept}(a,b)]. RGS70 (QAT~\#151) has 70 data qubits on a line, 70 CZ layers (2,415 CZ gates) and 27 check qubits; its output is measured after single-qubit rotations into a non-Clifford ``face-state'' basis. The experiment produced 1,726 post-selected samples, 0.10\% of the shots, in 1,950~s, with a measured fidelity of 0.37(2). DCS64 (QAT~\#245) has 64 data qubits on a ring, 73 CZ layers (2,336 CZ gates), 12 check qubits and 314 T gates inside the circuit. The experiment produced 1,389 post-selected samples in 870~s, with a certified fidelity of at least 0.349 at 95\% confidence.

\paragraph*{What counts as a classical challenge.} The experimenters specified the task for RGS70: produce uncorrelated samples from the rotated graph state with linear XEB significantly above 0.092, in a time comparable to the experiment. For DCS64 the quantity reported by the experiment is a fidelity lower bound. We report both figures of merit for both instances where they are defined: the fidelity of the state from which our samples are drawn, and the linear XEB of the samples, $2^n\langle p(x)\rangle-1$, scored with ideal probabilities.

\section{Exact sampling of the undoped instance}
\label{sec:exact}

The undoped instance prepares $\ket\psi=\bigl(\bigotimes_jU_j\bigr)\ket G$, a graph state followed by single-qubit unitaries $U_j$. Its entanglement across a balanced cut is as large as it can be, 35 bits for RGS70, which is why matrix-product-state estimates are so high. The same property makes sampling easy.

Choose a bipartition into halves $A$ and $B$ whose cut rank is $n/2$ [Fig.~\ref{fig:concept}(c)]. The reduced state on $B$ is then maximally mixed, and it remains so under the local unitaries. Measuring $B$ in any product basis therefore gives uniformly random bits: the outcome $b$ can be drawn with a classical random-number generator, with no reference to the circuit. All the structure of the distribution sits in the conditional state of $A$,
\begin{equation}
 \ket{\psi_{A|b}}\ \propto\ U_A\sum_{x_A}(-1)^{q_A(x_A)}\,g_b(\Gamma_{AB}x_A)\,\ket{x_A},
 \label{eq:cond}
\end{equation}
where $q_A$ collects the edges inside $A$ and $g_b$ is the Walsh--Hadamard transform of a function on $B$ that depends on the outcome $b$ and on the edges inside $B$. This is a vector of $2^{n/2}$ amplitudes, obtained with one fast transform, one relabeling of indices by the invertible matrix $\Gamma_{AB}$, and $n/2$ single-qubit gates. Sampling $a$ from it completes one exact sample $(a,b)$ of the full distribution. The cost per sample is $O(n\,2^{n/2})$ operations and $2^{n/2}$ amplitudes of memory, for any graph with a full-rank balanced cut and any product measurement basis (Theorem~\ref{thm:exact} and Algorithm~\ref{alg:undoped} in Appendix~\ref{app:proofs}).

Three remarks put this in context. First, the method is the measurement-based picture~\cite{Raussendorf2001} read backwards: measuring half of a maximally entangled pair of registers steers the other half into a state that is cheap to write down. Neither the circuit depth nor the non-stabilizerness of the measurement basis enters. Second, full-rank balanced cuts are common. A random binary matrix is invertible with probability 0.29, and for RGS70 we found a cut of rank 35 within four random trials. Third, the cost is exponential in $n/2$ with a small prefactor. One conditional state takes 256~GiB at $n=70$ in single precision, 8~TiB at $n=80$ and 256~TiB at $n=90$; undoped instances of this family become impractical for GPU clusters somewhere between 80 and 90 qubits. A chain-rule sampler along an order of smaller cut rank reduces memory further at the price of more passes~\cite{QAT}.

\section{Certified sampling of the doped instance}
\label{sec:doped}

T gates inside the circuit destroy the graph-state structure, and the argument above no longer applies. We proceed in three steps.

\paragraph*{Amplitudes from an enlarged graph state.} Each T gate can be replaced by a gadget qubit prepared in $\ket0$, a CNOT from the data qubit to the gadget qubit, and a projection of the gadget qubit onto $\bra0+e^{i\pi/4}\bra1$~\cite{ZhouLeungChuang2000,BravyiKitaev2005,Renaud249,Kuyanov2026}. Before the projections, the circuit prepares a stabilizer state on $n+t=378$ qubits, hence a graph state up to local Clifford gates. An amplitude of the doped circuit is a sum over the $t$ gadget variables, which we carry out one vertex at a time along a linear order. The intermediate table has $2^{\rho}$ entries, with $\rho$ the cut rank between the vertices already summed and the rest. The cost is set by the largest cut rank along the order, the width $w$; the smallest width over all orders is the linear rank-width of the enlarged graph.

\paragraph*{Sampling.} We leave $k=20$ output bits open during the contraction, so that one pass returns the amplitudes of all $2^{20}$ completions of a given 44-bit prefix $y$. Each sample draws a fresh prefix uniformly, computes its block, and draws the remaining 20 bits exactly from it~\cite{Manabe2026}. If all prefixes were equally likely this would be exact. In general it replaces the true prefix probabilities $p_y$ by $1/D$ with $D=2^{44}$, and the samples are those of the state $\sum_yD^{-1/2}\ket y\ket{\hat\psi_y}$, where $\ket{\hat\psi_y}$ is the normalized conditional state. Its fidelity with the ideal state is
\begin{equation}
 F=\Bigl(\E_y\sqrt{D\,p_y}\Bigr)^2 ,
 \label{eq:fid}
\end{equation}
with $y$ uniform. For a state that looks random on 44 qubits, $Dp_y$ concentrates around one and $F$ is close to one; Eq.~\eqref{eq:fid} lets us measure this instead of assuming it.

\paragraph*{Certificate.} The expectation in Eq.~\eqref{eq:fid} is over uniform prefixes, which is the distribution the sampler itself uses. Before the run we fixed 64 of the 1,389 sample indices as audits. For each, an independent double-precision contraction recomputes the full block, and we evaluate the overlap $c_y$ between the reference block, scaled by $\sqrt D$, and the normalized single-precision vector that was actually used for sampling. Then $\sqrt F=\E_yc_y$, with numerical errors of the sampling vectors included (Proposition~\ref{prop:cert}); Algorithm~\ref{alg:doped} summarizes the full procedure, illustrated in Fig.~\ref{fig:concept}(d). Because $1-c_y$ is bounded, a Chernoff bound on its mean gives a lower bound on $F$ at any confidence level without any assumption on the distribution of $p_y$. The bound controls the total-variation distance through $\TV\le\sqrt{1-F}$.

A sharper statement about the distribution is available. The TV loss $L_y=\sum_z\bigl(r_y(z)-|\psi_y(z)|^2/\pi_y\bigr)_+$, which compares the conditional law $r_y$ of the implementation with the ideal block, has mean exactly equal to the total-variation distance (Theorem~\ref{thm:tv}). We derived it after the run and report its value separately in Sec.~\ref{sec:results}.

\section{Implementation}
\label{sec:impl}

\paragraph*{Hardware.} All runs used one node with eight NVIDIA H200 GPUs (140~GiB of memory each), CuPy 13.6 and CUDA 12.9. Samples are independent in both algorithms, so the GPUs work in parallel without communication between groups and the run time scales inversely with their number.

\paragraph*{RGS70.} A binary tableau~\cite{AaronsonGottesman2004} reduces the Clifford circuit to a graph state on 70 vertices with 1,150 edges; an independent calculation with Stim~\cite{r3-stim} confirms all 70 stabilizers. The conditional state of Eq.~\eqref{eq:cond} has $2^{35}$ single-precision complex amplitudes (256~GiB) and is split across two GPUs. Transforms on the low-order bits are fused into single kernels, those on the high-order bits exchange data between the two GPUs, and the index relabeling by $\Gamma_{AB}$ is absorbed into the transform so that no data are moved. Four groups of two GPUs run in parallel.

\paragraph*{DCS64.} We start from the contraction code of Ref.~\cite{Renaud249} and change two things. Tables are updated in place in a single buffer, which halves the memory, and a search over slicing variables trades table size against the number of slices that are summed coherently. The production plan uses four slices and has width 34, i.e.\ one 128~GiB single-precision buffer per GPU; the reference plan for audits uses eight slices in double precision. One block of $2^{20}$ amplitudes takes about 4.9~s. Each GPU runs one worker with a preassigned share of the samples, its own random-number stream and its preassigned audits. Every record stores the generator state, the prefix, the interval of the cumulative distribution that was hit and a hash of the block, so that all random choices can be replayed.

\paragraph*{Validation.} The RGS70 sampler agrees with state-vector simulation on 16 small graphs and on induced subgraphs of the target with up to 64 vertices. The DCS64 blocks agree with state vectors on 168 small circuits (relative error $3\times10^{-15}$ in double precision) and, on the full circuit, with an independent double-precision contraction to $5\times10^{-7}$. The probabilities of all classical samples were recomputed in double precision for scoring.

\section{Results}
\label{sec:results}

\begin{table}[tb]
\caption{Hardware and classical results. Classical runs used eight H200 GPUs. $^a$Fidelity of the graph state before the final rotations, measured by direct fidelity estimation; it equals that of the measured state to first order in the infidelity~\cite{QAT}. $^b$Lower bound at 95\% confidence; for the classical sampler it is the fidelity with the ideal state of the pure state whose output distribution the sampler reproduces (Proposition~\ref{prop:cert}). $^c$Two independent batches of 1,726 samples, pooled; the timed batch alone gives 1.096(35). $^d$Scored with the probabilities of Ref.~\cite{Renaud249}. $^e$End to end, including scheduling, start-up and double-precision scoring of every sample; the sampling itself takes less.}
\label{tab:results}
\begin{ruledtabular}
\begin{tabular}{lcccc}
 & \multicolumn{2}{c}{RGS70} & \multicolumn{2}{c}{DCS64} \\
 & hardware & classical & hardware & classical \\ \hline
Samples & 1,726 & 3,452 & 1,389 & 1,389 \\
Fidelity & 0.37(2)$^a$ & exact & $\ge0.349^b$ & $\ge0.908^b$ \\
Linear XEB & 0.466(31) & 1.025(24)$^c$ & 0.390(34)$^d$ & 1.009(38) \\
Run time (s) & 1,950 & 1,265$^e$ & 870 & 864 \\
\end{tabular}
\end{ruledtabular}
\end{table}

\begin{figure}[tb]
\includegraphics[width=\columnwidth]{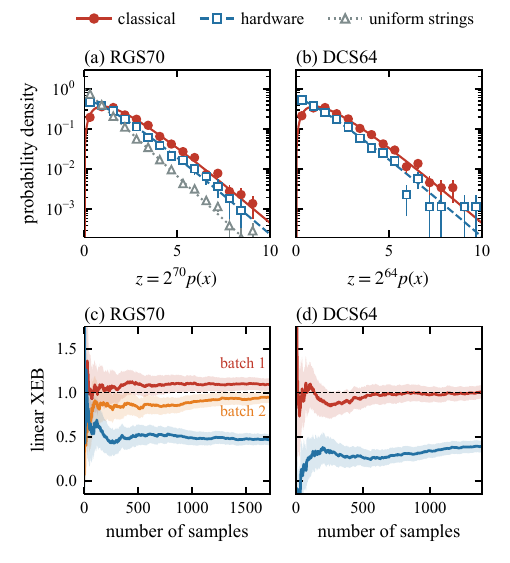}
\caption{Classical and hardware samples scored with ideal probabilities, for RGS70 (left) and DCS64 (right). (a,b) Distribution of the scaled probability $z=2^np(x)$. Solid lines: $ze^{-z}$, expected for an ideal sampler~\cite{Boixo2018}. Dashed lines: the white-noise mixture $Fze^{-z}+(1-F)e^{-z}$~\cite{Dalzell2024} with $F$ equal to the linear XEB of the hardware samples, 0.466 and 0.390. In (a), triangles and the dotted line $e^{-z}$ refer to 16,442 uniformly random bit strings. (c,d) Linear XEB as a function of the number of samples, with 95\% bands; the two classical batches of RGS70 are shown separately.}
\label{fig:results}
\end{figure}

\paragraph*{RGS70.} We generated two batches of 1,726 samples with independent seeds fixed in advance. Their linear XEB values are 1.096(35) and 0.953(33), and 1.025(24) together, consistent with the value 1 of an ideal sampler; no bit string occurs twice. The hardware samples give 0.466(31) in the same scoring. Figure~\ref{fig:results}(a) shows that the scaled probabilities of our samples follow $ze^{-z}$, while those of uniformly random strings follow $e^{-z}$ and the hardware samples a mixture of the two.

\paragraph*{DCS64.} The 64 audits have a mean overlap loss $1-c_y$ of $1.9\times10^{-4}$. The resulting bound is $\sqrt F\ge0.953$, so that $F\ge0.908$ at 95\% confidence, against 0.349 for the experiment. With only 64 audits, the bound could not exceed 0.911 even for a perfect sampler; it reflects the number of audits rather than the quality of the blocks. Scored in double precision, the 1,389 samples have linear XEB 1.009(38), and the hardware samples 0.390(34) [Fig.~\ref{fig:results}(b,d)]. Applied to the same 64 audits, the TV loss of Sec.~\ref{sec:doped} bounds the distance between our output distribution and the ideal one by 0.048 at nominal 95\% confidence, compared with 0.30 from the fidelity bound. We chose this analysis after the run, so it does not replace the fidelity statement fixed before the run.

\paragraph*{Run time.} For RGS70 the first batch took 1,265~s on a single monotonic clock, from the submission of the job to the return of all logs. This includes scheduling, start-up, sampling on the eight GPUs and double-precision scoring of every sample, and corresponds to 5.9 GPU seconds per sample. The second batch was run to test reproducibility and was not timed end to end; its workers took between 1,041 and 1,051~s. For DCS64 the eight workers started within 2~ms of each other. The slowest took 852~s to generate its samples and 864~s from the start of its process to its last written sample, which includes loading the plan and compiling the GPU kernels; this is 5.0 GPU seconds per sample. The 64 audits add 86~s per GPU, and the double-precision scoring of all samples took 4.05 GPU hours; both verify the samples and are not needed to produce them. The DCS64 figure excludes queueing and container start-up, so we regard the classical and quantum run times of this instance as equal within the uncertainty. In both cases doubling the number of GPUs would halve the time.

\paragraph*{Limitations.} Sampling uses single-precision arithmetic. For RGS70 the algorithm is exact in exact arithmetic, and the implementation is checked by double-precision rescoring but carries no rigorous error bound. For DCS64 the certificate includes the numerical error of the sampling vectors, but it is conditional on the double-precision reference blocks and on the pseudorandom prefixes behaving as independent uniform draws.

\section{Why the doped instance is tractable}
\label{sec:mechanism}

\begin{figure*}[tb]
\includegraphics[width=\textwidth]{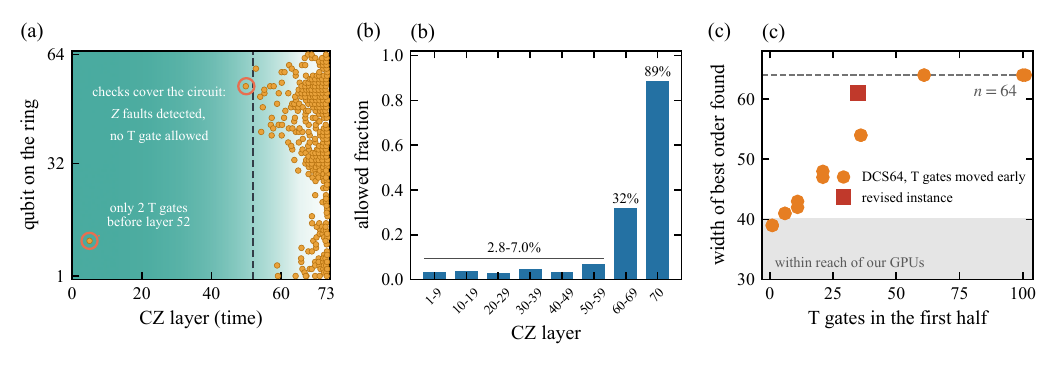}
\caption{Where T gates can go, and what early T gates cost. (a) Positions of the 314 T gates of DCS64 in the published circuit (points; circles mark the two before layer 52, dashed line). The shading indicates check coverage schematically. (b) Fraction of positions at which a T gate is compatible with all checks, by CZ layer, for the 70-qubit doped instance (QAT~\#228) with 27 checks, computed from the published circuit with its check qubits. (c) Width of the best linear order found in a 60~s search for DCS64 when some of its 314 T gates are moved to random positions in the first half of the circuit (circles, two seeds each; horizontal axis: number of T gates in the first half), and for the revised instance (square). Ordering the gadget qubits by time always gives width at most $n=64$ (dashed), and larger search values are plotted at 64. The shaded region is accessible to our implementation.}
\label{fig:mech}
\end{figure*}

\paragraph*{Where T gates can go.} A Pauli fault $P$ at some point in the circuit flips a check if it anticommutes with that check propagated backwards to the same point. A rotation about $Z$ on qubit $q$ leaves all checks invariant if every back-propagated check acts on $q$ as $I$ or $Z$; this is also the condition that a $Z$ fault on $q$ escapes detection~\cite{Martiel2025}. The allowed positions are therefore the places where the checks are blind to $Z$ faults. A check measured at the end of the circuit spreads as it is propagated backwards, and after a few layers it acts on a given qubit as $X$, $Y$ or $Z$ essentially at random. What matters locally is the space spanned by the restrictions of all checks to that qubit: if it has dimension 0, 1 or 2, a T gate is allowed with probability $1$, $1/3$ or $0$ under a random local frame (Proposition~\ref{r2:localrank}). Far from the output the dimension is almost always 2.

Figure~\ref{fig:mech}(b) shows the result for the 70-qubit doped instance of the same family, for which the circuit with its 27 check qubits is public. In CZ layers 1--59 only 2.8--7.0\% of the positions are allowed, and the experimenters used nearly all of them; the fraction rises to 32\% in layers 60--69 and 89\% in the final layer, layer 70. In DCS64, 278 of the 314 T gates lie in the last ten layers, and only one in the first 50.

\paragraph*{Late T gates are cheap.} A T gate close to the output can only influence the qubits in its forward light cone, and the gadget qubit that replaces it attaches to the enlarged graph near the boundary. We make this quantitative by cutting the circuit in time into an early region and a tail. Let the early axis rank $\kappa_E$ be the rank of the rotation axes of the T gates before the cut, propagated back to the input, that is, the number of independent early T gates; let the early entanglement $w_{\rm pre}$ be the largest entanglement, in bits, of the Clifford circuit before the cut across the cuts of a spatial sweep; and let $\gamma_{\rm tail}$ count the gates after the cut that cross those cuts. Then the enlarged graph has a linear order of width
\begin{equation}
 w\ \le\ \min\{\,n,\ \kappa_E+w_{\rm pre}+1+\gamma_{\rm tail}\,\}
 \label{eq:width}
\end{equation}
(Theorem~\ref{thm:geometry}). The T gates after the cut do not appear: they enlarge the graph, and the run time grows linearly with their number, but they do not increase the size of the tables. The early T gates enter only through the rank of their axes, and since every compatible T gate sits at an allowed position, $\kappa_E$ is at most the number $U_Z$ of early allowed positions. If single $Z$ faults in the early region are detected at a uniform rate $1-\epsilon$, then $U_Z$ is at most $\epsilon$ times the number of early positions. Better detection thus means less room for early T gates and a smaller width.

For DCS64 with the cut after layer 52, $\kappa_E=2$, $w_{\rm pre}=31$ and $\gamma_{\rm tail}=22$, and Eq.~\eqref{eq:width} gives 56. Most of this bound comes from the Clifford circuit before the cut; the two early T gates contribute only $\kappa_E=2$. Using the actual entanglement of the early region and of the tail as an operator, and optimizing over the cut and the origin of the sweep, lowers the bound to 43 (Theorem~\ref{r2:choi}, attained at the cut after layer 62). The best order known has width 36~\cite{Renaud249}, and we prove by exhaustive enumeration on a 28-vertex subgraph that no order has width below 10, so that $10\le w_{\min}\le36$ (Proposition~\ref{r2:finiteproof}).

\paragraph*{Early T gates are expensive.} Figure~\ref{fig:mech}(c) tests the complementary statement. We keep the 314 T gates of DCS64 but move some of them to random positions in the first half of the circuit, which ignores the checks. The width of the best order we find grows by about 0.4 per moved gate, from 39 to 54 for 36 early T gates, and reaches the trivial bound $n$ beyond about 60. Our implementation handles widths up to about 40, which this test exceeds at six early T gates. The revised instance posted by the experimenters has 35 T gates in its first 35 layers, with $\kappa_E=34$ instead of 1, and width 61 in our search; it is out of reach of the methods used here.

\begin{figure}[tb]
\includegraphics[width=\columnwidth]{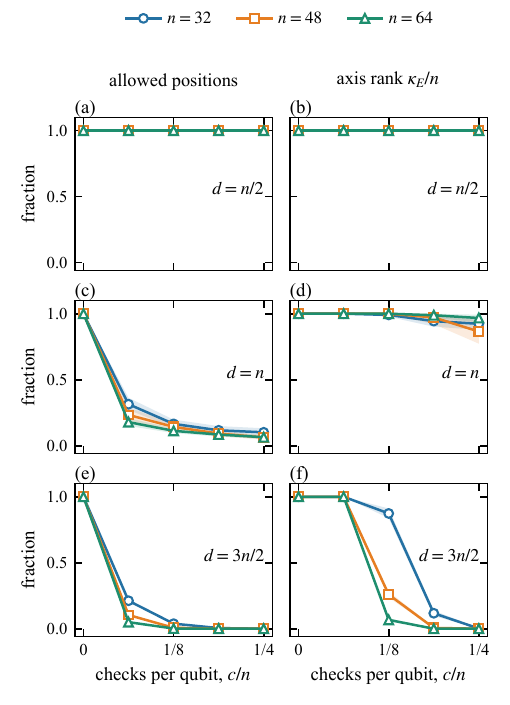}
\caption{Ensembles of random checked circuits with $n$ qubits and $c$ checks, for depths $d=n/2$ (a,b), $d=n$ (c,d) and $d=3n/2$ (e,f). Left: fraction of allowed positions in the first half of the circuit. Right: early axis rank $\kappa_E$ of the first half, divided by $n$. Points are means over 20 circuits; bands are pointwise 95\% bootstrap intervals.}
\label{fig:ensemble}
\end{figure}

\paragraph*{An ensemble of checked circuits.} To see how these quantities behave beyond single instances, we generated random circuits of the same type with $n=32$, 48 and 64 qubits and depths $d=n/2$, $n$ and $3n/2$, added up to $c=n/4$ checks with the experimenters' public tool~\cite{paulice}, and measured, for the first half of each circuit, the fraction of allowed positions and the early axis rank $\kappa_E$ (20 circuits per size and depth; Fig.~\ref{fig:ensemble} and Appendix~\ref{sec:r3-ensemble}). Three regimes appear. In shallow circuits, $d=n/2$, the checks do not reach the first half and every position is allowed. At $d=n$ the checks remove most positions, leaving 6\% at $n=64$ and $c=16$, but the few that remain still have axes of nearly full rank, $\kappa_E/n=0.97$. At $d=3n/2$ and $c=n/4$ the early region is closed: no position is allowed in 59 of 60 circuits. The rank, which is what enters Eq.~\eqref{eq:width}, thus drops abruptly rather than in proportion to the number of allowed positions, and only once the circuit is deep enough for the checks to cover the early region. DCS64, with $d=1.14\,n$, 12 checks and $\kappa_E=1$ in its first half, lies on the closed side. These numbers describe one generator with a fixed search budget, which differs in details from the one used for the experiments; a larger budget removes more positions without changing the rank.

\paragraph*{What does not follow.} It is tempting to summarize these observations as a law: more checks, fewer allowed positions, easier simulation, and conversely a weaker certificate for harder circuits. Each step fails in general.

(i) The early entanglement $w_{\rm pre}$ in Eq.~\eqref{eq:width} refers to the same spatial sweep as the other terms. It cannot be replaced by the width of the Clifford circuit optimized separately, even when no early position admits a T gate: there are circuits with one check, $U_Z=0$, optimized Clifford width one and at most eight layers after the cut whose enlarged graph has unbounded linear rank-width (Theorems~\ref{thm:counter} and~\ref{r2:zeroU}). These circuits are easy to simulate by other means, so they limit the inequality, not classical simulation.

(ii) The number of checks does not determine the number of allowed positions. Globally independent checks can have the same restriction to a qubit, and for the circuits and T gates of DCS64 and of the revised instance we exhibit different valid check circuits with different numbers of allowed positions (Sec.~S5 of the Supplemental Material~\cite{SM}). We could not reconstruct the experimental check circuits from the public data.

(iii) Many allowed positions do not force a loss of certified fidelity; there are doped circuits with arbitrarily many compatible T gates and no loss at all. Conversely, with few checks there exist compatible placements of large width (Theorem~S3).

(iv) A large width in the enlarged-graph representation is not a lower bound for other classical methods.

\section{Post-selection and the asymptotic comparison}
\label{sec:asymptotic}

Our samplers scale exponentially, as $2^{n/2}$ for undoped and $2^{w}$ for doped circuits, and for an ideal quantum computer the task is believed to be hard in the limit of large $n$. The experiments, however, pay for their fidelity with post-selection: RGS70 keeps 0.10\% of its shots. One might expect the acceptance rate to fall as the probability that no fault occurs, $e^{-\alpha nd}$ for $n$ qubits and depth $d$, which for $d\propto n$ would eventually cost more than $2^{n/2}$. This expectation is wrong. For independent faults and a syndrome of $r$ bits the acceptance rate is at least $2^{-r}$, because combinations of faults cancel in the syndrome (Appendix~\ref{app:theory}). With a number of checks proportional to $n$, post-selection costs at most $2^{O(n)}$; what decays with circuit volume is the fidelity of the accepted shots. A meaningful asymptotic comparison must therefore be made at fixed fidelity, where polynomial-time classical algorithms are known for sampling from noisy random circuits~\cite{Aharonov2023} and for estimating expectation values of noisy circuits~\cite{Schuster2024} and error mitigation has exponential overhead~\cite{Quek2024}, and not through the acceptance rate.

\section{Conclusion}

Both instances put forward to demonstrate verifiable quantum advantage with graph states can, at their present size, be sampled classically with higher quality than the hardware and in comparable or shorter time. The classical samples carry their own guarantee: they are exact for RGS70, and for DCS64 their fidelity is certified by audits fixed before the run. Classical challenges to verifiable experiments can thus be held to the same standard as the experiments themselves, a certified fidelity rather than a benchmark score alone.

More important than the two instances is the reason they fall. In both, the structure that makes the fidelity measurable is the structure that makes sampling easy. In the undoped instance, the Clifford structure that allows direct fidelity estimation also makes half of the output random. In the doped instance, the requirement that T gates be invisible to the checks pushes them to the end of the circuit, where they add little to the classical cost. Verification in this form competes with hardness for the same resource: non-Clifford gates early in the circuit.

This gives future experiments a concrete target. What matters is the early axis rank and the early entanglement. Both can be computed efficiently from the circuit before it is run, and both are better guides to classical hardness than the total number of T gates or the estimated cost of generic tensor-network contraction. Whether early T gates can be made plentiful without weakening the certificate is the question on which this route to verifiable advantage now depends. Our results do not answer it, and they do not suggest that the answer must be negative: the general trade-offs one might conjecture are false, and in our ensembles the early region is closed to T gates only when the circuit is somewhat deeper than it is wide. Comparisons in the limit of large circuits should be made at fixed fidelity, since post-selection itself is not what becomes prohibitive (Sec.~\ref{sec:asymptotic}).

On the classical side the samplers presented here are not optimal; tree-like contraction orders~\cite{GrayKourtis2021,Kuyanov2026}, stabilizer decompositions of the early T gates~\cite{BravyiGosset2016,Bravyi2019} and approximate methods with certificates are natural next steps. A first test is already available. Within a week of our classical submission for DCS64, the experimenters posted a revised instance (QAT~\#264) that, in their words, addresses this classical pressure: it places 35 T gates in the first half of the circuit, raising the early axis rank from 1 to 34, and the best order we find has width 61, beyond the methods used here. It also uses 7 checks instead of 12 and has a certified fidelity of 0.162 instead of 0.349. This is in line with the mechanism above, although two instances with different checked circuits do not define a trade-off curve. Whether this instance, or a successor that combines many early T gates with a strong certificate, withstands classical attack is the natural next question. Such an open cycle of claim and challenge, with circuits, samples and classical results all public, is how the boundary of quantum advantage is best mapped.

\begin{acknowledgments}
We thank the reviewers and contributors of the Quantum Advantage Tracker for making circuits, samples and probabilities public, and K.~Renaud for releasing the contraction code on which the DCS64 computation builds.
\end{acknowledgments}

{\raggedright \paragraph*{Data availability.} Samples, code, audit data and verifiers are available at \href{https://github.com/whyihaveyou/rgs70-classical-challenge}{github.com/\allowbreak whyihaveyou/\allowbreak rgs70-classical-challenge} and \href{https://github.com/whyihaveyou/dcs245-classical-sampling}{github.com/\allowbreak whyihaveyou/\allowbreak dcs245-classical-sampling}, and are archived on Zenodo (release v1.0.0) under DOIs \href{https://doi.org/10.5281/zenodo.24497053}{10.5281/zenodo.24497053} and \href{https://doi.org/10.5281/zenodo.24497055}{10.5281/zenodo.24497055}.\par}

\clearpage
\onecolumngrid
\appendix
\section{Exact conditional sampling and the fidelity certificate}
\label{app:proofs}

\begin{theorem}[Exact sampling across a full-rank cut]\label{thm:exact}
Let $\ket{\psi}=(\bigotimes_jU_j)\ket{G}$ with arbitrary single-qubit unitaries $U_j$, and let $A\cup B$ be a bipartition of the $n$ vertices with $|A|=|B|=n/2$ and $\rank\Gamma_{AB}=n/2$. Then the computational-basis marginal of $\ket\psi$ on $B$ is uniform, and exact samples from $|\langle x|\psi\rangle|^2$ can be drawn with $O(n\,2^{n/2})$ operations and $O(2^{n/2})$ memory per sample.
\end{theorem}
\begin{proof}
The reduced state of $\ket G$ on $B$ has rank $2^{\rank\Gamma_{AB}}=2^{n/2}$ with a flat spectrum~\cite{VandenNest2007}, so it is maximally mixed; local unitaries on $B$ preserve it, and $p(b)=2^{-n/2}$ for every $b$. For fixed $b$, write $q_A,q_B$ for the quadratic forms of $\Gamma_{AA},\Gamma_{BB}$ and
\[
 g_b(s)=\sum_{x_B}(-1)^{s\cdot x_B+q_B(x_B)}\prod_{j\in B}\langle b_j|U_j|x_j\rangle ,
\]
a Walsh--Hadamard transform over $n/2$ bits of a product-times-quadratic-phase function. The unnormalized conditional state on $A$ is $U_A\ket{\phi_b}$ with $\langle x_A|\phi_b\rangle=2^{-n/2}(-1)^{q_A(x_A)}g_b(\Gamma_{AB}x_A)$. Since $\Gamma_{AB}$ is invertible over $\F$, $x_A\mapsto\Gamma_{AB}x_A$ is a permutation of the $2^{n/2}$ indices. Computing $g_b$, permuting, and applying the $n/2$ single-qubit unitaries of $U_A$ each take $O(n\,2^{n/2})$ operations on one vector. Drawing $b$ uniformly and then $a$ from the normalized conditional distribution gives a sample from $p(a,b)=2^{-n/2}\,p(a|b)$.
\end{proof}
\begin{algorithm}[t]
\caption{Exact conditional sampling of an undoped instance}\label{alg:undoped}
\begin{algorithmic}[1]
\Require graph $G$ with adjacency matrix $\Gamma$, single-qubit unitaries $U_j$, number of samples $S$
\State find a balanced bipartition $A\cup B$ with $\rank\Gamma_{AB}=n/2$ by random trials
\For{$s=1,\dots,S$}
  \State draw $b\in\{0,1\}^{n/2}$ uniformly at random \Comment{marginal on $B$ is uniform}
  \State compute $g_b$ by one Walsh--Hadamard transform over the $n/2$ bits of $B$
  \State form $\phi_b(x_A)\propto(-1)^{q_A(x_A)}g_b(\Gamma_{AB}x_A)$ by relabeling indices
  \State apply the single-qubit unitaries of $U_A$ to obtain $|\psi_{A|b}\rangle$
  \State draw $a$ with probability $|\langle a|\psi_{A|b}\rangle|^2/\||\psi_{A|b}\rangle\|^2$ and output $(a,b)$
\EndFor
\end{algorithmic}
\end{algorithm}

Algorithm~\ref{alg:undoped} summarizes the sampler. A uniformly random square matrix over $\F$ is invertible with probability $\prod_{k\ge1}(1-2^{-k})\approx0.289$, so for random graphs a few random balanced cuts suffice.

\begin{proposition}[Fidelity of uniform-prefix sampling]\label{prop:cert}
Let $\psi_y\in\mathbb C^{2^k}$ be the unnormalized conditional vector of prefix $y\in\{0,1\}^{n-k}$, $\|\psi_y\|^2=p_y$, and let the sampler draw $y$ uniformly and the suffix from $|a_y|^2/\|a_y\|^2$ for some vectors $a_y\neq0$. With $D=2^{n-k}$ and
$c_y=|\ip{\sqrt D\,\psi_y}{a_y}|/\|a_y\|$,
the output distribution is the Born distribution of a pure state $\ket{\tilde\psi}$ with $|\langle\psi|\tilde\psi\rangle|=\E_y c_y$. For exact blocks, $F=(\E_y\sqrt{Dp_y})^2$.
\end{proposition}
\begin{proof}
Take $\ket{\tilde\psi}=\sum_yD^{-1/2}\ket y\otimes e^{i\theta_y}a_y/\|a_y\|$, with the phases $\theta_y$ chosen so that each $\ip{\psi_y}{e^{i\theta_y}a_y}$ is real and non-negative. Its Born distribution is that of the sampler, and $\langle\psi|\tilde\psi\rangle=\sum_yD^{-1/2}|\ip{\psi_y}{a_y}|/\|a_y\|=\E_yc_y$.
\end{proof}
\begin{algorithm}[t]
\caption{Certified uniform-prefix sampling of a doped instance}\label{alg:doped}
\begin{algorithmic}[1]
\Require doped circuit, linear order of the enlarged graph, prefix length $n-k$, samples $S$, audit set $\mathcal M\subset\{1,\dots,S\}$ fixed before the run, confidence $1-\alpha$
\For{$s=1,\dots,S$}
  \State draw a prefix $y\in\{0,1\}^{n-k}$ uniformly at random
  \State contract the enlarged graph in single precision, leaving the last $k$ output bits open, to obtain the block $a_y\in\mathbb C^{2^k}$
  \State draw the suffix $z$ with probability $|a_y(z)|^2/\|a_y\|^2$ and output $(y,z)$
  \If{$s\in\mathcal M$}
    \State recompute the block $\psi_y$ in double precision and record $c_y=|\langle\sqrt D\,\psi_y,a_y\rangle|/\|a_y\|$
  \EndIf
\EndFor
\State from the mean of $\ell_y=1-\min(c_y,1)$ over $\mathcal M$, compute the KL upper bound $u$ at confidence $1-\alpha$
\State \Return the samples and the certificate $F\ge(1-u)^2$
\end{algorithmic}
\end{algorithm}

Algorithm~\ref{alg:doped} summarizes the certified sampler. The overlap loss $\ell_y=1-\min(c_y,1)$ lies in $[0,1]$ and $\sqrt F\ge1-\E_y\ell_y$. For $m$ audited prefixes drawn independently and uniformly, with mean loss $\bar\ell$, the one-sided Kullback--Leibler Chernoff bound gives $\E\ell\le u$ with confidence $1-\alpha$, where $u\ge\bar\ell$ solves $m\,\kl(\bar\ell\,\|\,u)=\ln(1/\alpha)$~\cite{hoeffding}. For DCS64, $m=64$, $\bar\ell=1.94\times10^{-4}$ and $\alpha=0.05$ give $u=0.0469$, so $\sqrt F\ge0.9531$ and $F\ge0.9083$. Even with $\bar\ell=0$ the bound cannot exceed $\alpha^{2/m}=0.9106$; the number of audits, not the quality of the blocks, limits the certificate. As stated in Sec.~\ref{sec:results}, the bound treats the double-precision reference blocks as exact and the pseudorandom prefixes as independent uniform draws.

\section{Width bounds, counterexample and distributional audit}
\label{app:theory}

The appendices use the following notation. RGS70 and DCS64 are the instances posted as QAT~\#151 and \#245; \#228 and \#264 are the doped 70-qubit instance and the revised 64-qubit instance. The enlarged graph of Sec.~\ref{sec:doped}, before the gadget qubits are projected, is called the enlarged graph $G_U$, and $\lrw(G_U)$ is its linear rank-width, the smallest width over all linear orders. An allowed position is a slot at which a single $Z$ fault has zero syndrome, and $U_Z(E)$ is the number of allowed positions in the early region $E$.

\subsection{Model and compatibility}

An $n$-qubit logical circuit starts in $\ket{0^n}$ and contains Clifford gates
and rotations $R_{Z_q}(\theta)=e^{-i\theta Z_q}$; note that this convention has
no factor $1/2$, so that a T gate is $e^{i\pi/8}R_Z(\pi/8)$. The circuit has no
intermediate measurements or resets. When the checks are implemented
explicitly, their check qubits are part of the circuit; $N$ denotes the total number
of qubits and $n$ the number of data qubits.

Let $M_1,\ldots,M_c$ be final commuting Pauli syndrome observables with known
ideal outcomes. At a wire slot $s$, their backward images through the Clifford
skeleton are $Q_j(s)=C_{>s}^{\dagger}M_jC_{>s}$. A Pauli fault $P$ has syndrome
bit $[P,Q_j(s)]_{\rm sp}$, where the binary symplectic bracket is one for
anticommutation and zero for commutation.

\begin{lemma}[Local compatibility~\cite{Martiel2025}]
A single $T_q$ insertion preserves all skeleton check trajectories exactly when
$[Z_q,Q_j(s)]_{\rm sp}=0$ for every $j$. These are precisely the slots at which a
single $Z_q$ fault has zero syndrome. Any collection of individually compatible
insertions preserves the same trajectories.
\end{lemma}
\begin{proof}
Since $T_q=aI+bZ_q$ with $b\ne0$, it commutes with a Pauli $Q$ exactly when
$Z_q$ does. An anticommuting Pauli is changed by $T_q$ conjugation. A $Z_q$ fault
changes its sign exactly in the anticommuting case. Backward induction through
all insertions proves the last assertion.
\end{proof}

This rule and its fault interpretation are due to
Refs.~\cite{Martiel2025,Martiel2026,delfosse}. We apply it, as the experiments
do, to gate-by-gate compatible insertions; preserving the syndrome of the whole
circuit is a weaker condition, as $TT^{\dagger}=I$ shows.

The number of independent checks alone does not determine how many positions
are allowed: the checks $Z_1,\ldots,Z_c$ allow every $Z$ rotation, whereas the
single check $X_1\cdots X_n$ detects every single-qubit $Z$ fault. If the $c$
checks were conjugated by a uniform random Clifford on $N$ qubits, a fixed
nonzero Pauli would commute with all of them with probability
\begin{equation}
 \frac{2^{2N-c}-1}{2^{2N}-1},
 \label{eq:mixing}
\end{equation}
because the Clifford group acts transitively on nonzero Pauli vectors and the
commutant of the checks has dimension $2N-c$. Local circuits with optimized
checks need not be close to this limit (Appendix~\ref{app:local}).

\subsection{A rank and geometry bound for the enlarged graph}

Choose a temporal cut $\tau$. Let $C_{\rm pre}$ be the Clifford skeleton before
that cut, and let $C_j$ be the skeleton immediately before early rotation $j$.
Write
\begin{equation}
 P_j=C_j^{\dagger}Z_{q_j}C_j=\eta_jX(x_j)Z(z_j),\qquad
 V_E=\spanF\{x_j:j\text{ early}\},\qquad \kappa_E=\dim V_E.
 \label{eq:kappa}
\end{equation}
The phase $\eta_j$ makes $P_j$ Hermitian, and $\kappa_E\le\min(n,t_E)$.

Replace each rotation by a fresh $\ket{+}$ gadget $a$ and $\CZ_{a,q}$.
Contracting it with $\cos\theta\bra0-i\sin\theta\bra1$ recovers
$2^{-1/2}R_{Z_q}(\theta)$. Before these contractions the enlarged state is a
stabilizer state, and $G_U$ denotes a graph state locally Clifford equivalent to
it. A Hadamard on the gadget turns this into the injection of
Sec.~\ref{sec:doped}, with the gadget in $\ket0$ and a CNOT from the data qubit.

Fix one spatial order $q_1,\ldots,q_n$, let $A_q=\{q_1,\ldots,q_q\}$, and define
\begin{equation}
 w_q=\log_2\SR_{A_q:\bar A_q}(C_{\rm pre}\ket{0^n}),\qquad
 w_{\rm pre}=\max_q w_q.
 \label{eq:wpre}
\end{equation}
The order is kept fixed when accounting for the tail. For a graph state the
logarithmic Schmidt rank across $A$ is the binary rank of
$\Gamma_{A,\bar A}$: after removing phases internal to either side, its
coefficient matrix is the character matrix $(-1)^{a^T\Gamma b}$, with
$2^{\rank\Gamma}$ orthogonal distinct rows. This is the standard connection
between graph width and tensor descriptions~\cite{VandenNest2007,vidal}.

\begin{lemma}[Early part of the enlarged graph]
Order early gadgets chronologically. At the cut $\tau$ the enlarged state obeys
\begin{align}
 \log_2\SR_{E_{\le j}:(E_{>j},\mathcal D)}&\le\kappa_E,\label{eq:earlycut}\\
 \log_2\SR_{(E,A_q):\bar A_q}&\le\kappa_E+w_q.\label{eq:spatialcut}
\end{align}
\end{lemma}
\begin{proof}
Commuting the skeleton Cliffords to the end, without reordering the controlled
Paulis, gives
\[
 \ket{\mathcal R_E}=C_{\rm pre}
 \prod_{j=t_E}^{1}\mathrm C_{a_j}(P_j)
 \ket{0^n}_{\mathcal D}\ket{+^{t_E}}_E.
\]
Every Pauli changes a data basis label by $x_j$ and a phase, so data is supported
on $V_E$ before $C_{\rm pre}$. A binary basis change maps $V_E$ into $\kappa_E$
coordinates. The other coordinates remain zero; their $Z$ factors act trivially.
Fresh gadgets thus interact sequentially with a memory of dimension
$2^{\kappa_E}$. Future interactions and $C_{\rm pre}$ act on the opposite side
of an early-gadget cut, proving~\eqref{eq:earlycut}.

Alternatively, collect by final data label:
\[
 \ket{\mathcal R_E}=\sum_{x\in V_E}\ket{v_x}_E C_{\rm pre}\ket{x}.
\]
Each $C_{\rm pre}\ket{x}$ is a tensor-product Pauli translate of
$C_{\rm pre}\ket0$ and hence has the same Schmidt ranks. Subadditivity of
matrix rank over at most $2^{\kappa_E}$ terms, with all early gadgets on one
side, proves~\eqref{eq:spatialcut}. No commutativity between the $P_j$ is required.
\end{proof}

For CZ or CNOT tail gates define $h_q$ as the number crossing $A_q:\bar A_q$.
For a wire time $s$, define $g_{q,s}$ as the number of tail gates before $s$
crossing $A_q:\bar A_q$, plus those after $s$ crossing
$A_{q-1}:\bar A_{q-1}$; this is a separator that moves in time, not a fixed
spatial cut. Let $\gamma_{\rm tail}=\max\{h_q,g_{q,s}\}$, maximized over $q$ and
over all wire times $s$.

\begin{theorem}[Geometry-aware injection bound]\label{thm:geometry}
There is an explicit full vertex order $\sigma_\tau$ satisfying
\begin{align}
 w_{\sigma_\tau}(G_U)
 &\le\max\left\{\kappa_E,\ \max_q(\kappa_E+w_q+h_q),\
              \max_{q,s}(\kappa_E+w_q+1+g_{q,s})\right\}\label{eq:refined}\\
 &\le\kappa_E+w_{\rm pre}+1+\gamma_{\rm tail}.\label{eq:coarse}
\end{align}
Consequently $\lrw(G_U)$ is no larger. The order places early gadgets first
chronologically, then, for each spatial column, its late gadgets chronologically
followed by its data output.
\end{theorem}
\begin{proof}
An internal early-gadget cut satisfies~\eqref{eq:earlycut}; all tail operations
remain on the other side. After a full spatial column,
\eqref{eq:spatialcut} gives the initial bound. Each crossing CZ or CNOT has
operator Schmidt rank two and at most doubles the Schmidt rank; local gadgets
do not cross this cut. This gives the middle term of~\eqref{eq:refined}.

For a cut inside column $q$, put $A_q$ on the left at $\tau$, follow that
separator until $s$, cross the one data wire $q_q$, and then follow the
$A_{q-1}$ separator to the output. This puts precisely the required earlier
gadgets of that column on the left, and its later gadgets and data output on the
right. The initial separator has dimension at most $2^{\kappa_E+w_q}$; the data
wire adds a factor two; each of the $g_{q,s}$ crossing gates adds another factor
two. Contracting each side yields the final term. Graph cut-rank equals the
logarithmic Schmidt rank of the enlarged state before the projections.
\end{proof}

For depth-$d$ nearest-neighbor matching layers in the natural physical wire
order, $\gamma_{\rm tail}\le d$ on an open chain and $\gamma_{\rm tail}\le2d$ on
a ring: at the switching layer the two candidate moving edges share one wire
and cannot both belong to the matching. A general two-qubit Clifford gate
contributes $\log_2\operatorname{OSR}(g)\le2$ in place of one. The tail may
contain arbitrarily many single-qubit rotations; they enlarge the graph but not
this width term. The run time also depends on the contraction
algorithm~\cite{Kuyanov2026,markov}.

There is also a universal bound $\lrw(G_U)\le n$: order all gadgets
chronologically, then the data outputs. Gadget cuts are mediated by an
$n$-qubit sequential memory, and cuts in the final data block have at most $n$
qubits remaining. Hence $\lrw(G_U)$ is at most the minimum of $n$ and
\eqref{eq:refined}.

\begin{corollary}[Detection coverage and the allowed span]
Let $\mathcal L_E$ be a fixed slot set covering every allowed early insertion.
Let $\mathcal A_E$ consist of its zero-$Z$-syndrome slots and put
$U_Z=|\mathcal A_E|$ and $R_E=\rank\{x_\ell:\ell\in\mathcal A_E\}$.
Then every gate-by-gate compatible doping on these slots satisfies
\begin{equation}
 \lrw(G_U)\le w_{\rm pre}+R_E+1+\gamma
             \le w_{\rm pre}+U_Z+1+\gamma_{\rm tail}.\label{eq:detection}
\end{equation}
If the uniformly weighted early-slot detection rate is at least $1-\epsilon$,
one may replace $U_Z$ by $\epsilon|\mathcal L_E|$.
\end{corollary}
\begin{proof}
All actual early axes lie in the allowed span, so $\kappa_E\le R_E\le U_Z$;
apply Theorem~\ref{thm:geometry}. Repeated rotations at the same slot do not
enlarge the span. The coverage statement is the definition of uniform-slot
detection rate.
\end{proof}
The detection rate here is the unweighted fraction of detected positions; with
unequal fault weights it bounds $U_Z$ only if every weight is bounded below.

\subsection{A shallow-tail counterexample and its pruning conditions}

\begin{theorem}[Optimizing the Clifford order separately is invalid]\label{thm:counter}
No dimension-independent function $h$ can make
\[
 \lrw(G_U)\le\lrw(G_{\rm Clifford,full})+t_E+h(d_{\rm tail})
\]
hold for all nearest-neighbor Clifford-plus-$T$ circuits. There is a family
with $t_E=0$, $d_{\rm tail}=4$, and $\lrw(G_{\rm Clifford,full})=1$, but
unbounded $\lrw(G_U)$. A variant places every T gate immediately after a native CZ
with $d_{\rm tail}=8$. Both survive the pruning rules of Ref.~\cite{Martiel2026},
which remove rotations with input- or output-diagonal axes and duplicate axes.
\end{theorem}
\begin{proof}
Let $L_m$ be an $m\times m$ grid and $H_m$ its row-snake Hamiltonian path [Fig.~S1(a,b)]. Use
that path as the physical wire order. The graph $G_0=L_m\setminus E(H_m)$
contains only disjoint vertical path segments, so $\lrw(G_0)=1$ for $m\ge2$.
Prepare $\ket{G_0}$ by a Clifford prefix; nearest-neighbor SWAP routing with
restored wire order suffices, with no restriction on prefix depth.

On every path edge $\{i,j\}$ implement
\[
 e^{-i\pi Z_iZ_j/8}=\CNOT_{i\to j}R_{Z_j}(\pi/8)\CNOT_{i\to j},
\]
using the two edge matchings in four two-qubit layers. Finish with $H$ on every
data qubit. Removing the rotations cancels the CNOT sandwiches, leaving a state
locally Clifford equivalent to $\ket{G_0}$. In the enlarged graph,
\[
 \CNOT_{i\to j}\CZ_{a,j}\CNOT_{i\to j}=\CZ_{a,i}\CZ_{a,j}.
\]
Thus each path edge is replaced by a degree-two gadget, while the $G_0$ edges
remain. Local complementation at each gadget followed by deletion restores
$L_m$. Local complementation preserves cut-rank; deletion cannot increase
linear rank-width. Hence $\lrw(G_U)\ge\lrw(L_m)$~\cite{oum,jelinek}.

For completeness, a self-contained divergence bound suffices. Any grid vertex
set of size between $m^2/3$ and $2m^2/3$ has at least $m/6$ crossing edges:
either at least $m/6$ rows are mixed, or there exist both a full row and an empty
row, forcing one vertical crossing in every column. A crossing graph of
maximum degree four has an induced matching of size at least its edge count
divided by $32$: greedily choose an edge and remove edges incident to either
endpoint's neighbors. The matching gives an identity submatrix of the binary
cut matrix. Every vertex order has a balanced prefix, so
$\lrw(L_m)\ge m/192$, which diverges.

Each rotation's input-frame and output-frame axes are $X_iX_j$. They are
distinct, mutually commuting and nondiagonal, so none of the three pruning
rules removes them. The final Hadamards are essential for this assertion.
To enforce the immediate-after-CZ slot rule, replace each central $T_j$ by
$\CZ_{ij}T_j\CZ_{ij}$, an identical operation, and decompose the surrounding
CNOTs into $H_j\CZ_{ij}H_j$. Each matching then uses four CZ layers, eight in
total. The enlarged graph and the pruning conclusions are unchanged.
\end{proof}

The family is easy to simulate by other means. It shows that the same-order
term $w_{\rm pre}$ in Theorem~\ref{thm:geometry} cannot be replaced by an
optimized Clifford width. The version of the inequality with $U_Z$ in place of
$t_E$ is treated in Appendix~\ref{app:local} (Theorem~\ref{r2:zeroU}).

\subsection{Direct distributional certificates for conditional sampling}

Let $P(y,z)=|\psi_y(z)|^2$ be a normalized target. Fix a proposal $\pi_y$ and
normalized conditional laws $r_y(z)$, including finite-grid sampling and any
fallback. Then $Q(y,z)=\pi_y r_y(z)$. On $\pi_y>0$, define
$v_y=\psi_y/\sqrt{\pi_y}$. For DCS64, $\pi_y=2^{-44}$ and each suffix block
has $2^{20}$ entries.

\begin{theorem}[A bounded direct-TV audit]\label{thm:tv}
The TV loss
\begin{equation}
 L_y=\sum_z\bigl(r_y(z)-|v_y(z)|^2\bigr)_+
 \label{eq:loss}
\end{equation}
obeys $0\le L_y\le1$ and $\E_{Y\sim\pi}L_Y=\TV(P,Q)$, including zero target
blocks and proposal support that omits target mass.
\end{theorem}
\begin{proof}
Each positive term is at most $r_y(z)$. Multiplying by $\pi_y$ and summing gives
$\sum_{y,z}(Q-P)_+$; omitted prefixes have no positive contribution. Equal
total masses make this exactly $\frac12\sum|Q-P|$.
\end{proof}
For a randomized conditional engine, averaging its realized loss upper-bounds
the TV of its averaged output law by convexity. With exact conditionals the
identity reduces to $\TV(P,Q)=\TV(p,\pi)$, already used for uniform-prefix
batched sampling~\cite{Manabe2026}. Here the audit accounts for the actual
approximate conditional law, without a Porter--Thomas assumption.

For $m$ independent audits retained in full, let $\bar L$ be their mean and
define $u_{\rm KL}$ by
\begin{equation}
 m\kl(\bar L\|u_{\rm KL})=\log(1/\alpha),\quad
 u_{\rm KL}\ge\bar L,\qquad
 \kl(a\|b)=a\log\frac ab+(1-a)\log\frac{1-a}{1-b}.
 \label{eq:kl}
\end{equation}
At $\bar L=0$ use $u_{\rm KL}=1-\alpha^{1/m}$; at $\bar L=1$ or $m=0$ use one.
Then $\Pr\{\TV(P,Q)>u_{\rm KL}\}\le\alpha$.
Indeed, for $U_i\in[0,1]$ with
$\E[U_i\mid\mathcal F_{i-1}]\ge\delta$ and $t<0$, the exponential chord bound
gives
\[
 \E[e^{tU_i}\mid\mathcal F_{i-1}]\le1-\delta+\delta e^t.
\]
Iterating, using exponential Markov, and minimizing over $t<0$ gives
$\Pr\{\bar U\le a\}\le e^{-m\kl(a\|\delta)}$ for $a<\delta$.
Inversion proves the claim for a number of audits fixed in
advance~\cite{hoeffding}.

\subsection{Application to RGS70 and DCS64}

For RGS70 (QAT~\#151), the reduced graph has a balanced $35+35$ partition with
binary cross rank 35. The reduced state of the graph state on $B$ is therefore
$I/2^{35}$, and local unitaries on either side preserve this, so a uniform
draw of the $B$ outcomes followed by an exact conditional draw on $A$ gives
exact samples.

For DCS64 (QAT~\#245), the logical circuit has 64 qubits, 73 CZ layers and 314
rotations. An exact binary calculation gives, at the cut after layer 52,
\[
 \kappa_E=2,\quad t_E=2,\quad w_{\rm pre}=31,\quad\gamma_{\rm tail}=22,
 \qquad \kappa_E+w_{\rm pre}+1+\gamma_{\rm tail}=56.
\]
The explicit order of the theorem has width 39 [Fig.~S1(c)], and the best order known has
width 36~\cite{Renaud249}; Appendix~\ref{r2:instance} gives sharper bounds. The
check circuit of DCS64 is not public, so its allowed positions cannot be
computed. For \#228, whose check circuit is public, all 468 T gates are at
allowed positions, and the input-$X$ axes of all allowed positions span all 70
dimensions despite the 27 checks.

The double-precision scores of all 1,389 classical samples give a linear XEB of
$1.009352\pm0.038115$ for the classical samples and $0.389817\pm0.033753$ for
the hardware. The reference contraction uses the same graph and contraction
code as the sampler, in double precision.

\begin{table}[t]
\centering
\caption{Analysis of the 64 audits of DCS64 fixed before the run. The
direct-TV and product bounds were chosen after the run. The last column splits
$\alpha$ equally over the three procedures (union bound).}
\label{tab:cert}
\begin{tabular}{lcc}
\toprule
Quantity & Individual $\alpha=.05$ & Family $\alpha_j=.05/3$\\
\midrule
Coherent clipped-KL $F$ lower bound & 0.908321 & 0.877558\\
Direct-TV KL upper bound & 0.047902 & 0.064219\\
Classical product $F$ lower bound & 0.910710 & 0.879975\\
\bottomrule
\end{tabular}
\end{table}

The analysis in Table~\ref{tab:cert} and Fig.~S1(d) uses only the 64 stored pairs of sampling and reference blocks.
The mean of $L_y$ over the audits is $3.89\times10^{-4}$. The fidelity route
gives $\TV\le\sqrt{1-F}\le0.303$; the direct audit is tighter mainly because it
avoids the square root, while classical and coherent overlaps are nearly
identical on these blocks.

\subsection{Post-selection acceptance is not the no-fault probability}

\begin{proposition}[Independent-fault syndrome formula]
Let independent fault indicators $e_j$ have probabilities $p_j$, and let the
zero-syndrome event be $He=0$ over $\F$. Replace redundant rows of $H$ by an
$r$-row basis. Then
\begin{equation}
 A=\Pr(He=0)=2^{-r}\sum_{u\in\F^r}
       \prod_j(1-2p_j)^{(H^Tu)_j}.
 \label{eq:accept}
\end{equation}
If all $p_j\le1/2$, then $A\ge2^{-r}$.
\end{proposition}
\begin{proof}
Expand $1_{He=0}=2^{-r}\sum_u(-1)^{u\cdot He}$ and use independence. When
$p_j\le1/2$ all terms are nonnegative, including the unit contribution $u=0$.
\end{proof}
This is a standard character/weight-enumerator identity~\cite{macwilliams}.
It applies to the independent Pauli--Lindblad binary-channel model with
$p_j=(1-e^{-2\lambda_j})/2$, $\lambda_j\ge0$, used in the check-scoring
model~\cite{Martiel2026}.

For one check detecting every individual fault, $H=(1,\ldots,1)$ and equal
$p\in(0,1/2)$ give
$A=[1+(1-2p)^M]/2\to1/2$, whereas the no-fault probability $(1-p)^M\to0$.
Detecting every single fault therefore does not imply exponential rejection of
all faulty runs. If $r(n)=O(n)$, the acceptance cannot decay as $e^{-\alpha n^2}$
for fixed $\alpha>0$. High acceptance does not mean high fidelity, however:
many accepted multiple faults may be harmful.

\section{Local detection and the allowed-position conjecture}
\label{app:local}

We use the notation of Appendix~\ref{app:theory}. Recall the width bound of Theorem~\ref{thm:geometry},
\begin{equation}
 \lrw(G_U)\le\min\{n,\kappa_E+w_{\rm pre}+1+\gamma_{\rm tail}\},
 \label{r2:oldbound}
\end{equation}
and that compatibility of every early T gate gives $\kappa_E\le U_Z(E)$, where $U_Z(E)$ counts the allowed positions among a fixed set $\mathcal L_E$ of early candidate positions that contains every actual insertion. This appendix asks what controls $U_Z(E)$ locally, and whether $U_Z(E)$ can replace $\kappa_E$ in Eq.~\eqref{r2:oldbound}.

\subsection{The local statistic}
\label{r2:local}

Ignore Pauli phases, let $W\le\F^{2N}$ be the span of the back-propagated checks at a slot, and let $\pi_q(W)\le\F^2$ be its restriction to site $q$. Write $\beta_q=\dim\pi_q(W)\in\{0,1,2\}$.

\begin{proposition}[Local restriction law]
\label{r2:localrank}
The number of undetected nonidentity local Pauli axes at $q$ is $2^{2-\beta_q}-1$. After a fresh uniform single-qubit Clifford conjugates all check restrictions at this site together,
\begin{equation}
 \Pr(Z_q\text{ undetected}\mid W)=f(\beta_q),\qquad
 (f(0),f(1),f(2))=(1,1/3,0).
 \label{r2:twirl}
\end{equation}
For independent such twirls at one time slice, $\E(U_Z\mid W)=N_0+N_1/3$, where $N_i=|\{q:\beta_q=i\}|$, and, for $N_1>0$,
\begin{equation}
 \Pr\{U_Z-N_0-N_1/3\ge t\mid W\}\le e^{-2t^2/N_1}.
\end{equation}
\end{proposition}
\begin{proof}
The symplectic form on $\F^2$ is nondegenerate, so $\pi_q(W)^\perp$ has dimension $2-\beta_q$. Excluding the identity gives the first count. A uniform Clifford rotates each nonzero axis uniformly among $X,Y,Z$ and rotates the entire subspace by the same map, proving Eq.~\eqref{r2:twirl}. At rank zero or two the outcome is deterministic. At the $N_1$ remaining sites independent fresh twirls give independent Bernoulli variables of mean $1/3$; Hoeffding's bound gives the displayed inequality.
\end{proof}

For example, the $k$ independent commuting checks $Z_1Z_2,\ldots,Z_1Z_{k+1}$ all cover site 1, but their local span has rank one. The twirled probability that $Z_1$ is undetected is $1/3$ for every $k$, rather than $2^{-k}$. Conversely, $X_1X_2$ and $Z_1Z_2$ commute globally but have local rank two, so $Z_1$ is always detected. Thus the proposed local survival factor $2^{-k}$ does not follow from the number of covering independent checks. In a globally uniform Clifford orientation the distribution of local ranks instead yields
\begin{equation}
 a_{N,c}=\frac{2^{2N-c}-1}{4^N-1}.
 \label{r2:global}
\end{equation}
Indeed, a fixed nonzero fault is mapped uniformly to the $4^N-1$ nonzero Paulis; the centralizer of $c$ independent checks contains $2^{2N-c}-1$ of them.

\subsection{The published checked circuit of \#228}

For the published 97-wire checked circuit associated with QAT \#228, we computed local restriction ranks using the candidate convention of Fig.~\ref{fig:mech}(b): data wires immediately following data-data CZ gates. Table~\ref{r2:ranktable} gives an exact decomposition of its allowed positions. At rank one, let $N_{1,Z}$ count local span $\{I,Z\}$; then $U_Z=N_0+N_{1,Z}$, whereas fresh independent local twirls have mean $N_0+N_1/3$.

\begin{table}[htbp]
\centering
\caption{Local rank decomposition for the \#228 candidate positions after data-data CZ gates. The last column is the mean under independent local twirls, Eq.~\eqref{r2:twirl}.}
\label{r2:ranktable}
\begin{tabular}{rrrrrrr}
\toprule
Layers & Slots & $N_0$ & $N_1$ & $N_2$ & $U_Z$ & $N_0+N_1/3$\\
\midrule
1--35 & 2414 & 23 & 195 & 2196 & 91 & 88\\
1--59 & 4070 & 33 & 382 & 3655 & 172 & 160.333\\
60--69 & 690 & 107 & 336 & 247 & 220 & 219\\
70 & 70 & 51 & 19 & 0 & 62 & 57.333\\
\bottomrule
\end{tabular}
\end{table}

The early suppression is mainly due to local rank-two restrictions; the late increase comes from rank-zero and rank-one restrictions. The twirl mean in the last column is close to the actual count, although the experimental checks are optimized for a noise model~\cite{Martiel2026} rather than drawn at random.

\subsection{The allowed-position conjecture}
\label{r2:conjecture}

Consider the proposed inequality
\begin{equation}
 \lrw(G_U)\stackrel{?}{\le}\lrw(G_{\rm Clifford,full})+U_Z(E)+h(d_{\rm tail}),
 \label{r2:conjectureeq}
\end{equation}
where the Clifford width is separately optimized and $h$ is independent of system size. It has two different answers depending on the slot universe.

\begin{proposition}[Input-inclusive degeneracy]
\label{r2:input}
If the early universe includes an input slot on each data wire before preparation, every deterministic Clifford Pauli check leaves all these $Z$ faults undetected. Hence $U_Z(E)\ge n$, and Eq.~\eqref{r2:conjectureeq} follows trivially from $\lrw(G_U)\le n$ for nonnegative $h$.
\end{proposition}
\begin{proof}
The backward image of a deterministic Pauli check stabilizes $\ket{0^n}$ up to its known sign. It is therefore $Z$-type and commutes with every input $Z_q$. The width bound follows by ordering gadgets chronologically, then all data outputs: the sequential data memory has dimension $2^n$. If check qubits are part of the simulated circuit, the memory count and input-slot universe must be expanded consistently.
\end{proof}

We therefore count only positions immediately after native CZ gates, as in Fig.~\ref{fig:mech}(b). In the following construction every such position in the early region is detected.

\begin{lemma}[Noncrossing matching preparation with no early allowed CZ slot]
\label{r2:matching}
For a noncrossing perfect matching $M$ on a line, one can prepare
\begin{equation}
 \ket{\Phi_M}=\prod_{\{a,b\}\in M}R_{Z_aZ_b}(\pi/4)\ket{+}^{\otimes n}
 \label{r2:matchingstate}
\end{equation}
using nearest-neighbor $R_X(\pm\pi/4)$ and $R_{ZZ}(\pm\pi/4)$. After the initial Hadamard preparation, every early native-CZ slot detects a local $Z$ fault under the propagated global parity check.
\end{lemma}
\begin{proof}
Let $Q=X^{\otimes n}$. Every displayed generator commutes with $Q$, and
\begin{equation}
 R_{Z_jZ_{j+1}}(\pi/4)=e^{-i\pi/4}S_jS_{j+1}\CZ_{j,j+1},
 \label{r2:native}
\end{equation}
with $S^\dagger$ for the negative angle. At the CZ micro-step, $X_jX_{j+1}$ becomes $Y_jY_{j+1}$ up to sign; the subsequent local $S$ gates return it to $X_jX_{j+1}$. Thus every local check factor remains $X$ or $Y$ at every micro-step, not only at macro-gate boundaries. All local $Z$ faults are detected.

To construct the state without inserting ordinary SWAP decompositions, use the Majoranas
\begin{equation}
 \gamma_{2j}=\Bigl(\prod_{q<j}X_q\Bigr)Z_j,\qquad
 \gamma_{2j+1}=\Bigl(\prod_{q<j}X_q\Bigr)Y_j.
\end{equation}
The nearest-neighbor rotations $W_\ell=e^{\pi\gamma_\ell\gamma_{\ell+1}/4}$ are $R_{X_j}(\pi/4)$ for $\ell=2j$ and $R_{Z_jZ_{j+1}}(\pi/4)$ for $\ell=2j+1$. For an edge $a<b$, put $A=2a+1$, $B=2b$, and $R=W_{B-1}\cdots W_{A+1}$. Conjugation by each $W_\ell$ sends $\gamma_\ell$ to $-\gamma_{\ell+1}$. Since $B-A-1$ is even,
\begin{equation}
 RW_AR^\dagger=e^{\pi\gamma_A\gamma_B/4}
 =R_{P_{a,b}}(\pi/4),\qquad
 P_{a,b}=Z_a\Bigl(\prod_{a<q<b}X_q\Bigr)Z_b.
\end{equation}
Process matching edges from inner to outer. Noncrossing implies that all sites strictly inside $(a,b)$ are paired among themselves and have already been prepared. Their product of $X$ operators stabilizes that state, since each completed pair has $X_uX_v=+1$. Therefore the string rotation acts on the current state exactly as $R_{Z_aZ_b}(\pi/4)$. Induction gives Eq.~\eqref{r2:matchingstate}. The circuit uses only the stated $R_X$ rotations and the native decomposition in Eq.~\eqref{r2:native}, with their inverses, establishing detection at every early native slot.
\end{proof}

\begin{theorem}[Counterexample without allowed early positions]
\label{r2:zeroU}
For every integer $h\ge1$, there is a logical circuit on $n_h=2^{h+2}-4$ data qubits with one deterministic Pauli check such that
\begin{equation}
 U_Z(E)=0,\quad \lrw(G_{\rm Clifford,full})=1,\quad
 d_{\rm tail,CZ}\le8,\quad \lrw(G_U)\ge\lceil h/2\rceil.
 \label{r2:zeroUresult}
\end{equation}
The early universe contains every native-CZ-after slot. All actual T gates are individually compatible, immediately follow a native CZ, and survive input-diagonal, output-diagonal, and duplicate-axis pruning. Thus Eq.~\eqref{r2:conjectureeq} is false under this slot convention for every size-independent $h(d_{\rm tail})$.
\end{theorem}
\begin{proof}
Take the complete binary tree $T_h$ of height $h$. Replace each vertex $v$ by a block of $\deg(v)$ ports. Order blocks in depth-first preorder; within a nonroot block put the parent port first and the child ports in reverse traversal order. Connect ports corresponding to original tree edges, obtaining a perfect matching $M_h$. Its arcs are nested or disjoint: a right-child arc encloses the left-child subtree, and a parent arc closes before its child's descendant arcs begin. Hence $M_h$ is noncrossing.

Join consecutive ports inside each block by local edges $J_h$. The graph $M_h\cup J_h$ is a subdivision of $T_h$: retain the middle port of each three-port internal block, one root port, and each leaf port, and suppress the remaining degree-two ports. It has $n_h=2|E(T_h)|=2^{h+2}-4$ vertices and $|J_h|=2^{h+1}-3$ local edges.

Prepare $\ket{\Phi_{M_h}}$ by Lemma~\ref{r2:matching}, and cut $E$ immediately afterwards. This gives $U_Z(E)=0$. Apply Eq.~(S18) for every edge of $J_h$. At each middle $T$, the check has no $X$ component on the target, so compatibility is individual. The local edges have two matching colors. For each color decompose the two CNOTs into $H_j\CZ_{ij}H_j$ and replace its middle $T_j$ by $\CZ_{ij}T_j\CZ_{ij}=T_j$. This uses four CZ layers per color and places $T$ immediately after CZ. The extra pair commutes with the $Z$ insertion and its controlled-Z injection; it changes neither compatibility nor the enlarged graph. Finally apply Hadamards on every data wire, converting the preserved $Q$ into the deterministic output check $Z^{\otimes n_h}$.

With all T gates deleted the sandwiches cancel. Equation~\eqref{r2:native} makes the output locally Clifford equivalent to a matching, with optimized width one. With gadgets retained, Eq.~(S19) shows that each $J_h$ edge is subdivided once more. The enlarged graph is locally equivalent to a subdivision of $T_h$. Suppressing a degree-two tree vertex is local complementation followed by deletion, so $T_h$ is a vertex-minor. For forests, linear rank-width equals pathwidth, and $\operatorname{pw}(T_h)=\lceil h/2\rceil$~\cite{r2AdlerKante}. This proves Eq.~\eqref{r2:zeroUresult}.

Every $J_h$ edge joins distinct matching pairs, so $Z_iZ_j$ has zero expectation in the prefix state. Its input-frame Pauli therefore cannot be diagonal on $\ket{0^n}$. Final Hadamards turn its output axis into $X_iX_j$. The effective axes commute and are distinct, so none is removed by the stated pruning tests. Finally, the sum of matching arc lengths is $h2^{h+1}$; the preparation in Lemma~\ref{r2:matching} uses $(4h-2)2^h+2$ CZs sequentially. The preparation depth of the prefix grows with $h$.
\end{proof}

The family does not contradict Eq.~\eqref{r2:oldbound}, which uses the prefix width in the \emph{same} physical order, and it is easy to simulate: in the Majorana basis the data evolution is fermionic Gaussian~\cite{r2JozsaMiyake}, and the enlarged graph is a tree, of rank-width one.

\section{Upper and lower bounds on the width of DCS64}
\label{r2:instance}

\subsection{Retaining the actual prefix and tail ranks}

Fix the spatial order $q_1,\ldots,q_n$ and let $A_q=\{q_1,\ldots,q_q\}$. Write $\mathcal A_E$ for the early gadgets, and $\ket{R_E}$ for the prefix injection state before any magic-state contractions. Define
\begin{align}
 e_q&=\log_2\SR_{(\mathcal A_E,A_q):\overline{A_q}}(\ket{R_E}),\\
 a_E&=\max_j\log_2\SR_{\mathcal A_{E,\le j}:(\mathcal A_{E,>j},\mathcal D)}(\ket{R_E}),
 \label{r2:prefixranks}
\end{align}
with the early gadgets in chronological order. The proof of the lemma on the early part of the enlarged graph in Appendix~\ref{app:theory} gives $e_q\le\kappa_E+w_q$ and $a_E\le\kappa_E$, where $w_q$ is the Clifford-prefix entropy on the same cut.

Let $\sigma$ order all early gadgets chronologically, then the late gadgets on each wire chronologically followed by that data output. For a complete-column cut let $h_q$ count tail CZ/CNOTs crossing $A_q$. For a partial cut in column $q$ at time $s$, let $g_{q,s}$ count tail gates before $s$ crossing $A_q$, plus tail gates after $s$ crossing $A_{q-1}$. Using the actual prefix rank in the same tensor-separator proof gives
\begin{equation}
 w_\sigma\le\max\left\{a_E,\ \max_q(e_q+h_q),\
 \max_{\substack{\text{actual late}\\\text{gadget cuts }(q,s)}}(e_q+1+g_{q,s})\right\}.
 \label{r2:refinedgeometry}
\end{equation}
Indeed each prefix product term supplies one term, each crossed CZ/CNOT has operator Schmidt rank two, and the partial-column separator cuts one additional dimension-two data wire. Columns without a gadget need no invented partial cut.

To retain cancellations among tail Clifford gates, let $V_{\rm tail}:\mathcal D_{\rm in}\to(\mathcal D_{\rm out},\mathcal A_{\rm tail})$ be its injection isometry and introduce its normalized Choi state
\begin{equation}
 \ket{J_{\rm tail}}=2^{-n/2}\sum_x\ket{x}_R V_{\rm tail}\ket{x}.
\end{equation}
For an output cut of $\sigma$, put input references $R_{A_q}$ on the left. At a complete-column cut the left outputs are all completed columns. At a partial cut in column $q$, the left outputs are the completed $q-1$ columns and the already processed gadgets of column $q$; the current data output remains on the right. Denote the logarithmic Schmidt rank of this Choi partition by $\chi_{q,s}$, using $s=\infty$ for a complete column.

\begin{theorem}[Prefix--Choi composition bound]
\label{r2:choi}
For the declared order,
\begin{equation}
 w_\sigma(G_U)\le
 \max\left\{a_E,\ \max_{\text{actual output cuts }(q,s)}(e_q+\chi_{q,s})\right\}.
 \label{r2:choibound}
\end{equation}
In particular, $\chi_{q,\infty}\le h_q$ and $\chi_{q,s}\le1+g_{q,s}$, so this retains the information used in Eq.~\eqref{r2:refinedgeometry}.
\end{theorem}
\begin{proof}
Schmidt-decompose $\ket{R_E}$ across $(\mathcal A_E,A_q)\mid\overline{A_q}$ into $2^{e_q}$ product terms. Schmidt-decompose the Choi state across the specified reference/output partition into $2^{\chi_{q,s}}$ terms. Inverse vectorization of each side turns these into rectangular linear maps from its input side to its output side. Thus
\begin{equation}
 V_{\rm tail}=\sum_{\ell=1}^{2^{\chi_{q,s}}}L_\ell\otimes R_\ell,
\end{equation}
where input and output dimensions on a side need not coincide. Applying this to the prefix expansion produces at most $2^{e_q+\chi_{q,s}}$ product vectors across the final cut. Early-gadget cuts retain their prefix Schmidt ranks, since the entire tail acts on the other side. These cuts exhaust $\sigma$. Finally the same moving separator cuts no extra data wire for a complete column and one for a partial column; each crossing CZ/CNOT contributes at most a factor two, proving the comparisons.
\end{proof}

The unequal input/output partition in a partial column is essential: the input reference for $q$ is on the left while its data output is on the right. Ignoring this mismatch would lose the extra data-wire factor. Operator Schmidt decompositions and the Choi representation are standard~\cite{vidal,r2Tyson,r2Choi}. All ranks here are computed by exact binary elimination.

\subsection{Instance bounds and the subset certificate}

For the enlarged graph of DCS64 (QAT~\#245), with 64 data qubits and 314 gadget qubits, the finite searches in Table~\ref{r2:bounds} improve the analytic bounds. The original refined geometry expression is optimized over all 74 temporal cuts and 128 cyclic origins/directions. The Choi scan covers all 74 cuts in the natural order and 128 ring orders at nine selected cuts. The best Choi decomposition is at cut 62, origin 39, forward direction, with 30 early T gates and $\kappa_E=27$. One critical partial cut has $e_q=32$ and $\chi_{q,s}=11$. The actual width of that constructed order is 40; the order of Ref.~\cite{Renaud249} is better, at 36.

\begin{table}[htbp]
\centering
\caption{Upper bounds on the linear rank-width of the enlarged graph $G_U$ of DCS64. The chronological order gives 64.}
\label{r2:bounds}
\begin{tabular}{lr}
\toprule
Certificate & Upper bound\\
\midrule
Original coarse geometry at cut 52 & 56\\
Original $\kappa_E+w_q$ refined geometry, finite scan & 54\\
Actual prefix ranks, geometric tail & 44\\
Actual prefix ranks, Choi tail & 43\\
Earlier explicit natural spatial order & 39\\
Complete 378-vertex order of Ref.~\cite{Renaud249}, checked on every cut & 36\\
\bottomrule
\end{tabular}
\end{table}

For a genuine lower bound, take a specified 28-vertex induced subgraph $H$ of this enlarged graph. Write $\rho_H(S)=\rank\Gamma_{S,V(H)\setminus S}$ and define
\begin{equation}
 \Lambda(\varnothing)=0,\qquad
 \Lambda(S)=\max\{\rho_H(S),\min_{v\in S}\Lambda(S\setminus\{v\})\}.
 \label{r2:subset}
\end{equation}
Every cut rank in this recurrence is relative to the \emph{whole} graph $H$, even when $S$ is a proper subset.

\begin{proposition}[Finite integer certificate]
\label{r2:finiteproof}
Equation~\eqref{r2:subset} computes $\Lambda(V(H))=\lrw(H)$. The recorded exhaustive calculation for the specified 28-vertex subgraph gives 10, whereas the verified complete order gives 36. Consequently
\begin{equation}
 \boxed{10\le\lrw(G_U)\le36.}
 \label{r2:interval}
\end{equation}
\end{proposition}
\begin{proof}
Among permutations of $S$ ending at $v$, the previous prefixes are precisely those of a permutation of $S\setminus\{v\}$; the new last prefix has rank $\rho_H(S)$ independent of $v$. Minimizing over $v$ proves the recurrence by induction on $|S|$. It is a standard subset layout dynamic program, distinct from an exact algorithm for tree rank-width~\cite{r2OumExact}.

The integer implementation enumerates all $2^{28}=268435456$ subsets and obtains $\Lambda(V(H))=10$. A threshold-reachability recurrence also finds no complete order of width nine, and an explicit order attains ten. Restricting any ordering of $G_U$ to $H$ cannot increase any corresponding cut rank, because the cut matrix of $H$ is a submatrix of the original cut matrix. Therefore $\lrw(H)\le\lrw(G_U)$. Checking every cut of the complete 378-vertex order of Ref.~\cite{Renaud249} gives the upper endpoint 36.
\end{proof}

The lower endpoint is a computer-assisted proof. The enumeration takes about 100~s on a CPU; the subgraph, the code and the optimal order are included in the data release. The Choi construction was checked against dense calculations on 36 small circuits, and every output cut of the full enlarged graph satisfies Eq.~\eqref{r2:choibound}.

\section{Ensembles of checked random circuits}
\label{sec:r3-ensemble}

The structural bound depends on the independent early rotation axes, not on
the number of checks alone. We therefore generate an ensemble of checked
circuits with the experimenters' public check-search tool,
\texttt{qiskit-paulice} version 0.3. The exact settings used for the
experiments are not public, so the ensemble is of the same type but not
identical.

\subsection{Circuit ensemble and candidate positions}
The payload has $n\in\{32,48,64\}$ data qubits on a ring. Starting from
$|0\rangle^{\otimes n}$, we apply $H$ to every qubit, followed by $d$ alternating
nearest-neighbor CZ matching layers. After each matching, independently at
each site, we apply either $\sqrt X$ or $S\sqrt X$ ($\sqrt X$ first), with equal
probability. The depths are $d/n\in\{1/2,1,3/2\}$. As in
Ref.~\cite{Martiel2026}, a final layer of local Clifford gates turns the output
into a graph state: Stim's graph-state decomposition~\cite{r3-stim} gives
$C|0\rangle=L|G_0\rangle$, and we append $L^\dagger$, written in the gates $S$
and $H$, and verify the signed stabilizers against $|G_0\rangle$. This layer is
part of the circuit seen by the check search; it adds no entangling gates and
no candidate positions.

The candidate set $\mathcal L^{\mathrm{CZ}}$ contains the two data endpoints
immediately after every CZ of the payload, so $|\mathcal L^{\mathrm{CZ}}|=nd$;
positions inside the added check gates are excluded. The early region is
\begin{equation}
 E=\{\ell\in\mathcal L^{\mathrm{CZ}}:t_\ell\le\lfloor d/2\rfloor\},
 \qquad |E|=n\lfloor d/2\rfloor.
 \label{eq:r3-region}
\end{equation}
In the checked circuit, the CZ gates between data qubits agree exactly, and in
order, with those of the payload, and each candidate is placed immediately
after the corresponding gate. No routing or transpilation is performed, so each
candidate refers to the same logical position in the checked and the unchecked
circuit.

\subsection{Three distinct observables}
After $c$ checks have been added, propagate each final check-qubit
$Z_{a_j}$ backwards through the complete checked Clifford circuit. At candidate
$\ell$, on data wire $q_\ell$, the local $X/Y$ indicator is the syndrome bit
$s_{j\ell}$. Let $s_\ell\in\mathbb F_2^c$ collect these bits and define
\begin{equation}
 A_c(E)=\{\ell\in E:s_\ell=0\},\qquad
 U_Z^{\mathrm{(CZ)}}(E)=|A_c(E)|.
 \label{eq:r3-allowance}
\end{equation}
These syndromes use the checked circuit, while the rotation axes are computed
in the $n$-qubit payload. If $C_\ell$ is the payload prefix through the original CZ,
write
\begin{equation}
 \begin{aligned}
 C_\ell^\dagger Z_{q_\ell}C_\ell
   &=\eta_\ell X(x_\ell)Z(z_\ell),\\
 R_E&=\operatorname{rank}_{\mathbb F_2}\{x_\ell:\ell\in A_c(E)\}.
 \end{aligned}
 \label{eq:r3-rank}
\end{equation}
As in Eq.~\eqref{eq:kappa}, $R_E$ is the rank of the input-$X$ components.

\paragraph{Noiseless logical correspondence.}
For the all-zero input, the checked circuit has the form
$W_T=(H_A\otimes I_{\mathcal D})V_T(H_A\otimes I_{\mathcal D})$, where
$H_A=H^{\otimes c}$ and $V_T=\sum_b|b\rangle\langle b|_A\otimes V_{T,b}$ is block
diagonal in the computational basis of the check qubits, including diagonal phase
corrections. The controlled-Pauli structure of the checks gives
$V_{T,0}=e^{i\theta}U_T$, where $U_T$ is the bare payload with the retained
rotations. Each check qubit's preparation Hadamard precedes all its couplings and
its readout Hadamard follows all of them, with only diagonal check-qubit operations
in between; we verify these conditions on every generated circuit.

Compatibility guarantees that, in the output Clifford frame, each inserted
rotation commutes with every final $Z_{a_j}$. Their ordered product therefore
preserves the joint $+1$ eigenspace. Since the undoped checked circuit has
zero syndromes, the doped circuit has output
$|0^c\rangle_A\otimes|\phi_T\rangle_{\mathcal D}$. Undoing the final Hadamards
and comparing the all-zero check-qubit component gives
\[
 2^{-c/2}V_{T,0}|0^n\rangle=2^{-c/2}|\phi_T\rangle.
\]
Thus the data output equals $U_T|0^n\rangle$ up to phase, for all retained
rotations together.

We filter the full allowed candidate list before inserting gates. Repeatedly,
the first chronologically applicable boundary rule removes an input-diagonal
axis commuting with all earlier active axes, or an output-diagonal axis
commuting with all later active axes; input is tested first. After boundary
rules reach a fixed point, the first lexicographic pair of equal unsigned
axes with no intervening anticommuting active axis loses its later candidate,
and boundary filtering restarts. Output axes are computed with the full
payload. This filter follows the rules of Ref.~\cite{Martiel2026} and the public
implementation~\cite{paulice}. A T gate is then assigned to every retained
candidate. If $K_c$ is the retained set, define
\begin{equation}
 \kappa_E=\operatorname{rank}_{\mathbb F_2}
       \{x_\ell:\ell\in K_c\cap E\}.
 \label{eq:r3-kappa}
\end{equation}
Filtering is performed on the full circuit, before restricting to $E$.

\paragraph{Deterministic checks.}
For every realization,
\begin{equation}
 0\le\kappa_E\le R_E\le\min\{n,U_Z^{\mathrm{(CZ)}}(E)\}.
 \label{eq:r3-chain}
\end{equation}
Indeed, the retained columns are a subset of the allowed columns, and a
binary matrix has rank at most its row and column counts. Adding a check
preserves the old syndrome bits, so $A_{c+1}(E)\subseteq A_c(E)$ and both
$U_Z^{\mathrm{(CZ)}}(E)$ and $R_E$ are nonincreasing in $c$. The filtered
$\kappa_E$ need not be monotone, since filtering is applied anew to each list.
The ensemble measures how strongly these quantities are suppressed.

In terms of the local rank $\beta_\ell\in\{0,1,2\}$ of the check restrictions,
with $N_0$ rank-zero positions and $N_{1,Z}$ rank-one positions whose nonzero
direction is $Z$,
\begin{equation}
 U_Z^{\mathrm{(CZ)}}(E)=N_0(E)+N_{1,Z}(E).
 \label{eq:r3-local-rank}
\end{equation}
Independent local Clifford twirls would instead give the mean $N_0+N_1/3$,
which we record for comparison. A rank-one $Z$ restriction detects local $X$
and $Y$ faults but allows the $Z$ rotation.

\subsection{Results and their structural interpretation}
The study contains 276 runs (180 main trajectories and 96 control runs) with 3,042 check searches, all of which succeeded.
Table~\ref{tab:r3-max-checks} reports the snapshots with the most checks; the full table for all 36 combinations is in the data release.
At $d=n$, reducing the number of allowed positions need not produce a proportional reduction of their input-$X$ rank.
At $d=n$ and $c=n/4$, the mean allowed fractions for $n=32,48,64$ are respectively 0.1013, 0.0622, 0.0642; the corresponding mean normalized ranks $R_E/n$ are 0.9234, 0.8646, 0.9672.
At $d=n/2$, all early positions remain allowed and both ranks equal $n$ in 240 of 240 nonzero-check snapshots. At $d=3n/2,c=n/4$, all three early observables vanish in 59 of 60 snapshots.
For one circuit chosen in advance ($n=d=64$, $c=16$, base index 0), increasing the window budget from 100 to 1,000 changes $U_Z^{\rm(CZ)}(E)$ from 511 to 77, while $\kappa_E$ remains 64. This illustrates why candidate count and rank must be measured separately.
In all 720 snapshots with checks, $R_E=\kappa_E$: the filter removes no early allowed position, although it removes positions elsewhere in the circuit.
\begin{table*}[tp]\centering\small
\caption{Mean primary observables at $c=n/4$, with pointwise 95\% bootstrap intervals. Each row contains 20 independent base trajectories.}\label{tab:r3-max-checks}
\begin{tabular}{rrccc}\toprule $n$ & $d/n$ & $U_Z^{\rm(CZ)}(E)/|E|$ & $R_E/n$ & $\kappa_E/n$ \\ \midrule
32 & 0.5 & 1.000 [1.000, 1.000] & 1.000 [1.000, 1.000] & 1.000 [1.000, 1.000] \\
32 & 1 & 0.101 [0.077, 0.132] & 0.923 [0.878, 0.964] & 0.923 [0.878, 0.964] \\
32 & 1.5 & 6.5e-05 [0.000, 2.0e-04] & 0.002 [0.000, 0.005] & 0.002 [0.000, 0.005] \\
48 & 0.5 & 1.000 [1.000, 1.000] & 1.000 [1.000, 1.000] & 1.000 [1.000, 1.000] \\
48 & 1 & 0.062 [0.047, 0.079] & 0.865 [0.773, 0.950] & 0.865 [0.773, 0.950] \\
48 & 1.5 & 0.000 [0.000, 0.000] & 0.000 [0.000, 0.000] & 0.000 [0.000, 0.000] \\
64 & 0.5 & 1.000 [1.000, 1.000] & 1.000 [1.000, 1.000] & 1.000 [1.000, 1.000] \\
64 & 1 & 0.064 [0.050, 0.079] & 0.967 [0.935, 0.992] & 0.967 [0.935, 0.992] \\
64 & 1.5 & 0.000 [0.000, 0.000] & 0.000 [0.000, 0.000] & 0.000 [0.000, 0.000] \\
\bottomrule\end{tabular}\end{table*}
At $d=n,c=n/4$, switching from the nested main procedure to fresh equally spaced target searches changes $\kappa_E/n$ by mean paired differences +0.0000, +0.0250, -0.0375 for $n=32,48,64$, respectively (five bases per size).
The mean within-base ranges of $\kappa_E/n$ across three checker seeds are 0.0063, 0.0000, 0.0031.
For the two budget-control bases per size, the mean paired differences in $\kappa_E/n$ (100 minus 1,000 windows) are +0.0156, +0.0000, +0.0000.

Controls on the check search, the paired sensitivity runs and the validation of the ensemble code are described in Sec.~S7 of the Supplemental Material.

\bibliography{refs}
\end{document}